\documentclass[aps,prb,reprint,groupedaddress]{revtex4-1}

\usepackage{graphicx}% Include figure files
\usepackage{epstopdf}
\usepackage{dcolumn}% Align table columns on decimal point
\usepackage{bm}% bold math
\usepackage{color}
\usepackage{natbib}
\usepackage{amsmath}
\usepackage{amssymb}

\begin{document}

\title{Origin of giant spin-lattice coupling and the suppression of ferroelectricity in EuTiO$_3$ from first principles}
%\title{Why EuTiO$_3$ is not a quantum paraelectric like SrTiO$_3$}
%\title{Unlike SrTiO$_3$, EuTiO$_3$ is not a quantum paraelectric}
%\title{Why EuTiO$_3$ is not SrTiO$_3$}
%\title{Giant hardening of a polar soft mode in EuTiO$_3$}

\author{Turan Birol and Craig J. Fennie}
\affiliation{School of Applied $\&$ Engineering Physics, Cornell University, Ithaca, NY 14853, USA}

\begin{abstract}
We elucidate the microscopic mechanism that causes a suppression of ferroelectricity  and an enhancement of octahedral rotations in  EuTiO$_3$ from first principles. We find that the hybridization of the rare earth Eu 4f states with the B-site Ti cation drives the system away from ferroelectricity. We also show that the magnetic order dependence of this hybridization is the dominant source of spin-phonon coupling in this material. Our results underline the importance of rare earth f electrons on the lattice dynamics and stability of these transition metal oxides.
\end{abstract}

\date{\today}

\maketitle

%%
%==================================================
\section{Introduction}
%==================================================

In the last ten years there has been an intense effort  to discover new  materials that display strong magnetoelectric coupling.  Such materials  could enable novel devices in which an electric-field control  magnetism.\cite{nan2008, martin2012, martin2012b, he2012} In this  pursuit  first-principles computational methods have played a key role by successfully  predicting new material realizations~\cite{cohen1993, franceschetti1999, spaldin2003, hafner2006, benedek2011, bousquet2011, birol2012, picozzi2012, yin2013, lee2010b, lee2011b}  even when the underlying microscopic mechanisms have not always been clear.\cite{giovannetti2012} Elucidating these mechanisms is important for both fundamental  understanding and also for guiding the search for new magnetoelectrics.  One example is the perovskite EuTiO$_3$.\cite{mcguire1966}

\begin{figure}[t]
  \begin{center}
    \includegraphics[width=0.85\hsize]{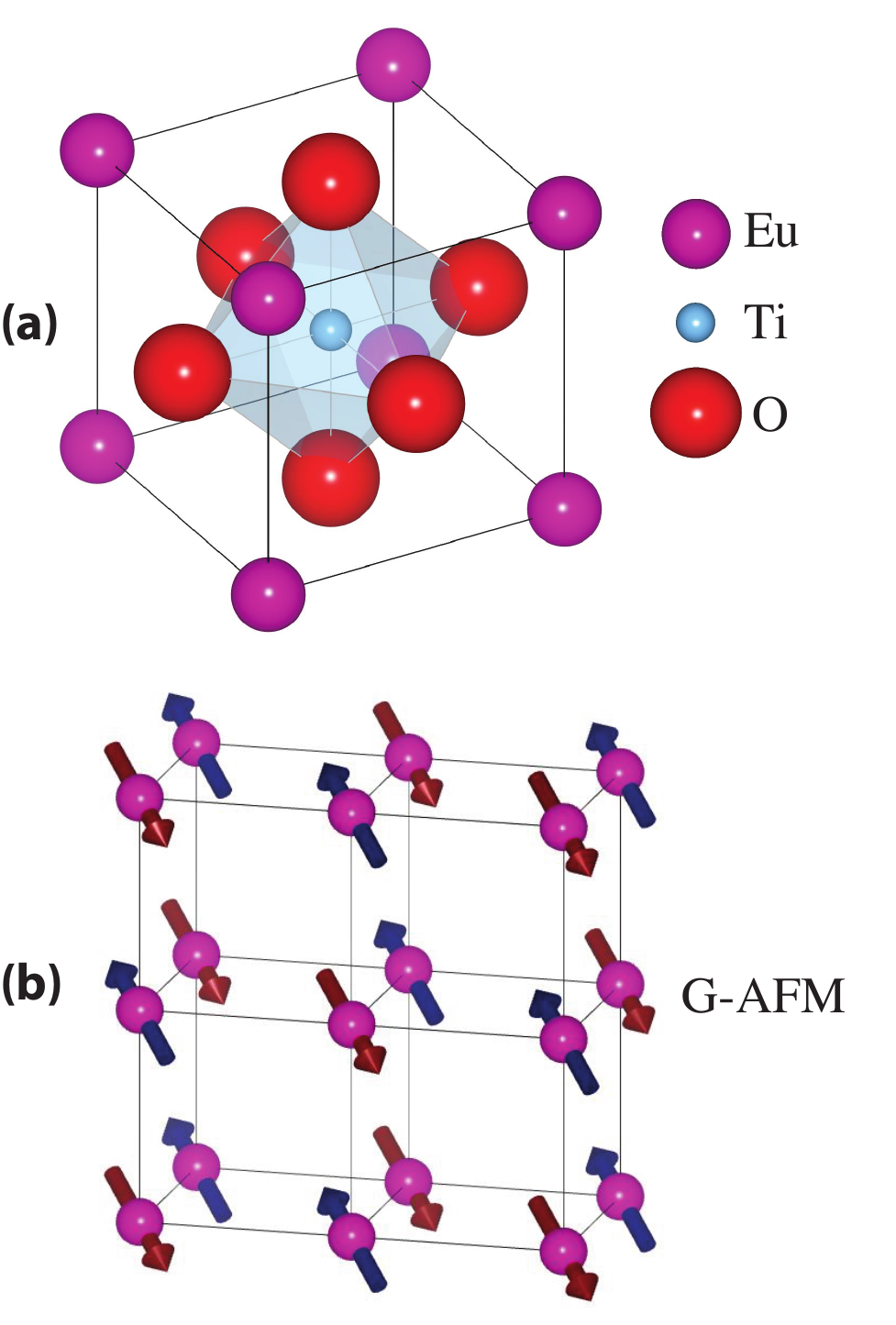}
  \end{center}
    \caption{(a) Crystal structure of perovskite EuTiO$_3$ in the cubic phase. (b) Sketch of G-type antiferromagnetic order. Arrows denote the direction of spins.}
    \label{fig:structure}
\end{figure}

  Bulk EuTiO$_3$, shown in Fig.~\ref{fig:structure}a, is a paraelectric antiferromagnet that displays a dielectric anomaly at the magnetic ordering temperature ($T_N\sim 5.3$ K).\cite{chien1974, katsufuji2001} Much more relevant  to the possibility of magnetoelectric phase control were the pioneering experiments of Katsufuji and Takagi,  which  showed that EuTiO$_3$ exhibits a magnetodielectric effect; at low temperatures, the dielectric constant  depends strongly on the magnitude of the external magnetic field.\cite{katsufuji2001} They suggested that this effect stemmed from spin-phonon coupling,\cite{lee2011a, hong2012} i.e., the dependence of the polar phonon frequencies on spin correlations. Subsequent first-principles studies~\cite{fennie2006, ranjan2007} and direct measurements of the phonon frequencies under magnetic field\cite{kamba2007, kamba2012a} have  largely confirmed this picture. 

%=====================================================
%=====================================================

Regarding the magnetoelectric properties of EuTiO$_3$, it was shown from first principles  how the underlying physics leading to the observation of spin-phonon coupling  can be exploited to enable control over the dielectric and the magnetic ground state.\cite{fennie2006} Epitaxial strain was proposed as the explicit control ``knob'' that tunes the antiferromagnetic-paraelectric ground state into a simultaneous   ferromagnetic-ferroelectric phase. Furthermore  it was argued that under increasing strain but before reaching this novel multiferroic phase, a giant magnetoelectric response  would occur in the vicinity of the strain-induced phase transitions due to phase competition.\cite{newnham1998, tokura2006} Subsequent experiments on epitaxially strained thin films have observed the strain induced ferromagnetic-ferroelectric phase~\cite{lee2010} and also the suppression of the antiferromagnetic  order by an external electric field,\cite{ryan2013} both of which are consistent with the original prediction, yet the giant magnetoelectric effect has yet to be observed (possibly due to a lack of high quality substrates that would provide the necessary  value of strain).

While it has been suggested that the physics of EuTiO$_3$ largely originates from a cation-mediated exchange mechanism,\cite{akamatsu2011, birol2012}  the microscopic mechanism of the spin-phonon coupling, and subsequently of the magnetoelectric phase control, is unknown. 
Here we ask  an important, but overlooked, question  whose answer makes these clear; \textit{Why isn't EuTiO$_3$ a ferroelectric in bulk}? 

\begin{figure}[t]
  \begin{center}
    \includegraphics[width=1.0\hsize]{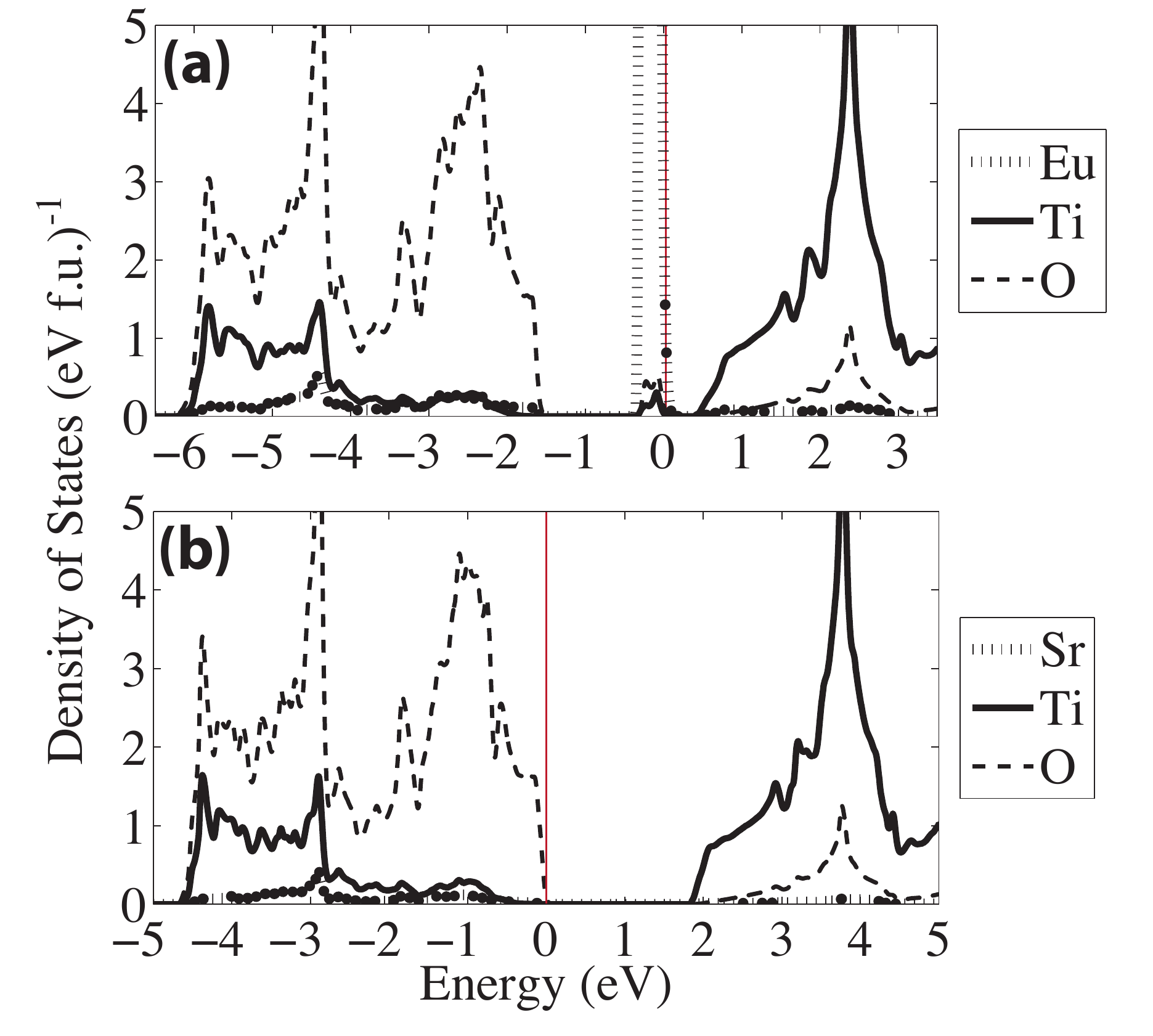}
  \end{center}
    \caption{(a) Density of states (DOS) of EuTiO$_3$, calculated from first-principles. (b) DOS of SrTiO$_3$,  calculated from first-principles. The zero points of energy in the DOS plots are aligned with the highest occupied level, and are also shown by the vertical red lines.}
    \label{fig:dos}
\end{figure}

Before we begin discussing our results we must make it clear in what sense we are ``surprised'' that EuTiO$_3$ is not ferroelectric in bulk. First, note the similarity to  the perovskite SrTiO$_3$: both compounds in the cubic phase have  almost identical lattice constants (to two significant figures), both have nominally Ti$^{4+}$ in an oxygen octahedral environment, and both have an A$^{2+}$ cation on the A-site. In fact these two perovskite compounds have very similar band structures. Both have a charge transfer gap -- between filled oxygen 2p states and  empty Ti d states -- of similar magnitude. The only major difference is the presence of narrow Eu 4f bands in EuTiO$_3$ (Fig. \ref{fig:dos}a and \ref{fig:dos}b). 
(The 4f character of the valence band is also experimentally observed.\cite{kolodiazhnyi2012})
These 4f electrons, however, are well localized and shielded by the 5s and 5p electrons. As a result, they are not expected to contribute significantly to chemical bonding. Because of these facts there is good reason to believe that the structural and dielectric properties of EuTiO$_3$ and SrTiO$_3$ should be quite similar.

Consistent with this conjecture is the observation that both compounds undergo a structural phase transition due to the softening of a zone-boundary, antiferrodistortive mode (corresponding to a rotation of the octahedra).\cite{rushchanskii2012, lee2010, yang2012, ellis2012} In EuTiO$_3$, however, this occurs at a much higher temperature ($\approx$100K for SrTiO$_3$, $\approx$300K for EuTiO$_3$).~\cite{fleury1968, allieta2012, ellis2012} Even more surprising  is the fact that SrTiO$_3$ displays a static dielectric constant of $\epsilon \sim 10^4$ at low temperature.\cite{muller1979, weaver1959} This huge dielectric constant has been explained  in a picture of a nominally unstable zone-center polar phonon mode being weakly stabilized by quantum fluctuations,~\cite{zhong1996, muller1979} and as such, is referred to as a quantum paraelectric. Indeed, first-principles calculations of the infrared-active (polar) phonons within Density Functional Theory, DFT,  have shown that at the experimental lattice constant, SrTiO$_3$ displays a weak ferroelectric instability. (All DFT studies of SrTiO$_3$ that we know of used the most common formulation of DFT, which is a static theory where fluctuations of the nuclei, quantum or thermal, are not considered). Quantum Monte Carlo studies of a first-principles parameterized effective Hamiltonian indeed show that this ferroelectric  state is suppressed by quantum fluctuations,\cite{zhong1996} consistent with a picture of SrTiO$_3$ being a quantum paraelectric.

In contrast, for EuTiO$_3$, first-principles calculations  of the polar phonons within DFT at the experimental cubic lattice constant have shown that all the polar modes are quite hard, with the softest mode being $\omega\sim +70$ cm$^{-1}$. It is therefore hard to imagine that EuTiO$_3$ is close to a ferroelectric phase transition. Consistent with this is the fact that the low temperature ($\sim$ 5K) dielectric constant of EuTiO$_3$ is two orders of magnitude smaller, $\epsilon\sim10^2$, than in SrTiO$_3$.~\cite{katsufuji2001} While there appears to be a ``rounding off'' of the dielectric constant below $\sim$ 10K in EuTiO$_3$, which people consider to be an observable effect of quantum fluctuations,  we stress that EuTiO$_3$ would remain a paraelectric even in their absence. This is clear from every DFT study ever conducted~\cite{fennie2006, rushchanskii2012, kamba2012a} and from the experimental determination of the Cochran fit to soft-mode frequency, which gives $\omega_{SO}\sim 75$ cm$^{-1}$ at zero temperature.\cite{kamba2007} Because of these facts we would not refer to EuTiO$_3$ as a quantum paraelectric.

Our question should therefore be understood within the following sense: given that lattice instabilities in A$^{2+}$TiO$_3$ perovskites tend to be controlled by volume (baring chemistry differences, e.g., the Pb$^{2+}$ lone-pair cation), and that  SrTiO$_3$ and EuTiO$_3$ have almost identical volumes in the cubic phase, what leads to the giant hardening of the soft polar mode is  EuTiO$_3$ compared with SrTiO$_3$? 
What seems to be the only possible explanation is that somehow the Eu 4f electrons have a giant effect on the lattice instabilities: dramatically decreasing (increasing) the tendency for EuTiO$_3$ to display a ferroelectric instability (rotational instability).

%=====================================================
%=====================================================

In this Article we discuss the answer to these questions, thereby providing a microscopic picture of spin-phonon, or more accurately spin-lattice, coupling in EuTiO$_3$ from first principles.  (Note, the physics of spin-phonon coupling that we are discussing is  in actuality a spin-lattice coupling, i.e., the effect of the magnetic order/correlations on the force constants, an inherently static quantity.\cite{birol2012})
%s
We explain our methods in Section \ref{sec:methods} and give a brief background on EuTiO$_3$ in Section \ref{sec:background}. 
In Section \ref{sec:U}, we explain the evolution of the polar soft-mode frequency under changing Hubbard-$U_{Eu}$. 
We then elucidate the key role played by the hybridization of the filled Eu 4f states with those of the nominally empty Ti d states in Section~\ref{hybrid} . 
In Section~\ref{ferro} we show how this leads to a giant hardening of the polar soft mode, subsequently driving EuTiO$_3$ away from ferroelectricity and  rendering  it a paraelectric with a small dielectric constant. 
In Section~\ref{spin_phonon} we  explore the magnetic order dependence of this hybridization and show how it is the dominant  cause of spin-lattice coupling. 
In Section~\ref{rotations} we argue that the much stronger oxygen octahedral rotations in EuTiO$_3$ compared to SrTiO$_3$ also originates from the hybridization of the Eu f electrons. 
Finally, we conclude with a summary in Section \ref{sec:summary}.
%

%==================================================
\section{Methods} \label{sec:methods}
%==================================================

First-principles calculations were performed within density functional theory using the PBE-GGA exchange-correlation functional\cite{PBE} and the Projector Augmented Wave method\cite{PAW1, PAW2} as implemented in VASP.\cite{VASP1, VASP2}   Because of the well-known deficiency of PBE-GGA  in describing the localized nature of f-electrons of e.g., Eu, the DFT+U formalism is used.\cite{DFTU:LDAUTYPE1_1, DFTU:LDAUTYPE1_2} The on-site exchange $J_{Eu}$ is kept fixed at 1.0 eV, while a Hubbard-$U_{Eu}=6.2$ eV was found to give the best fit to experiment (where we compared the ratio of the N$\acute{\rm e}$el to Curie temperatures  calculated within mean field theory).  The value of $U_{Eu}$, however, is often varied in our calculations in order to probe the physics of the system, as will be described. The cubic lattice constant is kept fixed to the experimental value of $a=3.90$ \AA.
Phonon frequencies are calculated using both Density Functional Perturbation Theory and frozen phonons  technique and no discrepancy is observed.  
We made extensive use of the Isotropy Software Package\cite{isotropy2007} and the Bilbao Crystallographic Server\cite{bilbao1,bilbao2,bilbao3,bilbao4}. Visualization of crystal structures are made using Vesta.\cite{vesta2008} Maximally Localized Wannier Functions (MLWF) are calculated using the Wannier90 code.\cite{WANNIER_REVIEW, wannier90}

%==================================================
\section{Results}
\label{results}
%==================================================

%==================================================
\subsection{Background} 
\label{sec:background}
%==================================================
%
The  cubic crystal structure of perovskite EuTiO$_3$ is shown in Fig.~\ref{fig:structure}a. Rotations of oxygen octahedra, which are known to exist in bulk EuTiO$_3$, were recently shown to have a strong effect on magnetism.~\cite{akamatsu2012, akamatsu2013, rushchanskii2012, yang2012, ryan2013, goian2012} In particular, they alter the magnetic exchange interactions in a way that strongly favors antiferromagnetism over ferromagnetism. Additionally, rotations tend to suppress the tendency towards ferroelectricity.\cite{yang2012, ryan2013} In the epitaxial  strain phase diagram of EuTiO$_3$ this results in an increase in the critical strain value necessary to induce a transition form the paraelectric-antiferromagnetic phase to the ferroelectric-ferromagnetic phase. For tensile strain, however, this increase in critical strain is almost canceled by the larger value of Hubbard-$U_{Eu}$, which lowers the critical strain, that is now necessary to give a reasonable fit of the magnetic exchange interactions in the presence of rotations to experiment (see Ref.~\onlinecite{ryan2013} for a complete discussion).

Magnetism in EuTiO$_3$ stems from the half filled 4f shell of the Europium cation, which have 7 electrons in a high-spin state. These well localized spins order in a collinear G-type antiferromagnetic fashion (Fig. \ref{fig:structure}b) so that the spin of each Eu cation is opposite to all of its nearest neighbors. As seen in the density of states (DOS) in Fig. \ref{fig:dos}a, there is a wide charge transfer gap between the the occupied O-p states and the conduction band that consists of unoccupied Ti-d states. The half-occupied Eu 4f states form narrow bands below the Fermi level in this charge transfer gap. There is very little hopping between the Eu-4f orbitals and the neighboring cations, since the radii of the 4f orbitals are much smaller than that of the 5s or 5p orbitals. This is the reason that the  N$\acute{\rm e}$el temperature is low, and the Eu-f bands are narrow. (The N$\acute{\rm e}$el temperature is further lowered because of the competition between ferromagnetic and antiferromagnetic exchange interactions.) The Hubbard-U applied on the Eu-f states ($U_{Eu}$) shifts the energy level of the narrow Eu-f bands, and hence determines the gap between them and the Ti d or O p bands. 
%and determines the location of the narrow Eu f bands.

%=====================================
\begin{figure}[t]
  \begin{center}
    \includegraphics[width=0.9\hsize]{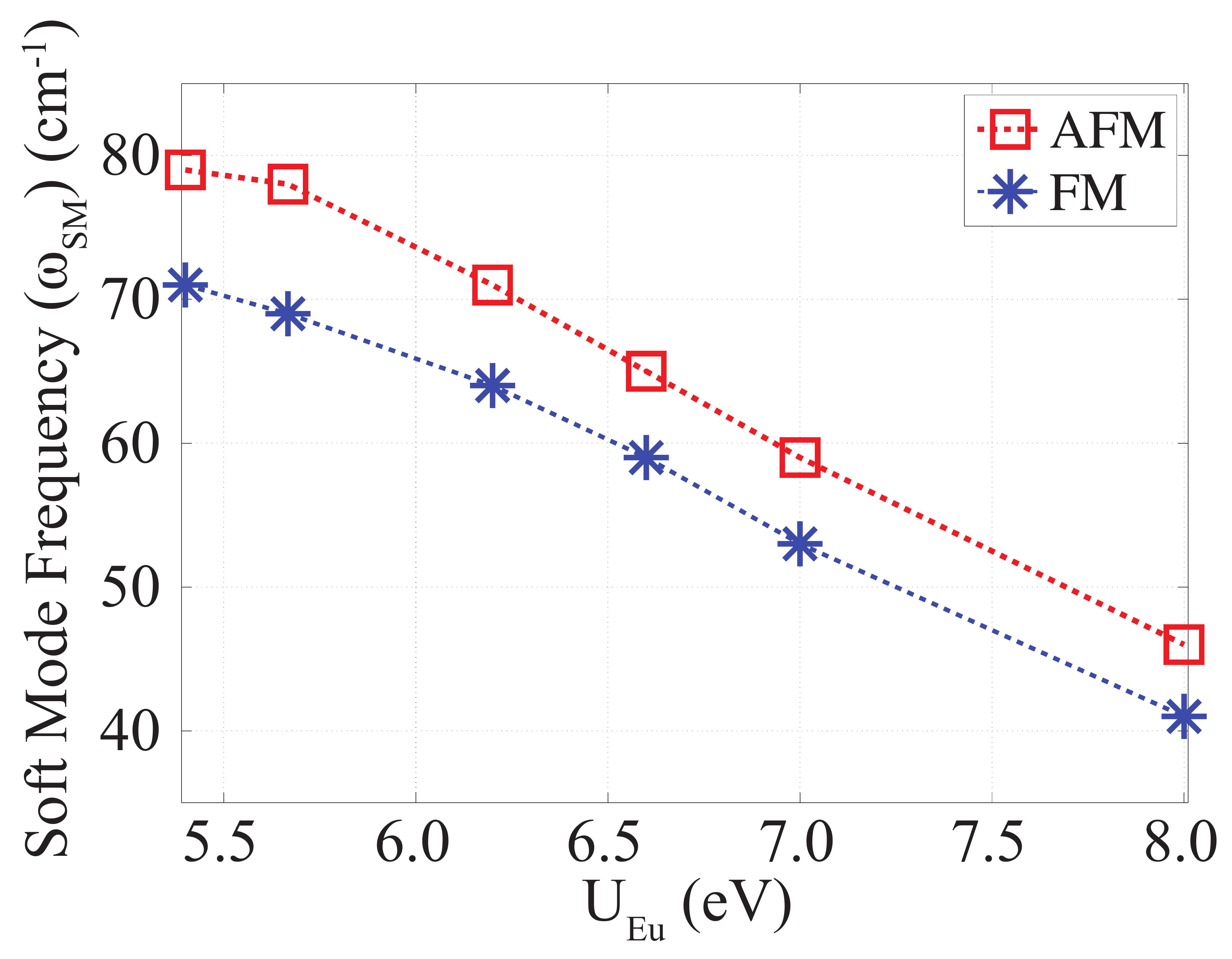}
  \end{center}
    \caption{Polar soft-mode frequency, $\omega_{SM}$, vs. the on-site interaction $U_{Eu}$ on Eu 4f orbitals. Red squares and blue asterisks denote the frequencies calculated in AFM and FM states respectively.}
    \label{fig:softmode}
\end{figure}
%=====================================

%=============================
\subsection{An Intriguing thought experiment: Polar mode frequency versus Hubbard-U}
\label{sec:U}
%==============================
%
To begin unraveling the mechanism behind the soft mode behavior of EuTiO$_3$, we perform a thought experiment where we calculate the frequencies,  $\omega_{SM}$,  of the polar phonons  of cubic (space group Pm$\bar{3}$m) EuTiO$_3$ from first principles, for several different values of the Hubbard-U (applied to the Eu-f states, $U_{Eu}$). This is plotted in Fig. \ref{fig:softmode}. 
Note that the frequency at a value of $U_{Eu}\sim6$ eV reproduces well the experimental value,  $\omega_{SM}\sim 75$ cm$^{-1}$, determined from a Cochran fit.\footnote{It is pleasing that the original DFT paper determined a value of U similar to this by comparing the calculated magnetic exchange parameters to experiment, and then predicted a $\omega_{SM}$ frequency remarkably close to the experiment.}

There are two clear trends in Fig.~\ref{fig:softmode}:  (i) as is now well-known, $\omega_{SM}$ is lower in the ferromagnetic (FM) state, which explains~\cite{fennie2006} an increase in the ionic contribution to the dielectric constant,~\cite{cochran1962} $\epsilon_{\rm ion}\sim 1/\omega_{SM}^2$, under external magnetic field, and (ii) $\omega_{SM}$ depends sensitively on $U_{Eu}$. With regards to the latter, notice how a relatively modest increase in $U_{Eu}$ greatly decreases $\omega_{SM}$. This is surprising as the polar soft-mode of EuTiO$_3$ is  driven by the off-centering of the Ti$^{4+}$ cation, i.e., B-site driven, in a second order Jahn-Teller like process~\cite{bersuker1978, cohen1992,rabe2007, bersuker2012} (we elaborate on this below), and therefore it is not expected to depend so sensitively on the energy of the Eu bands, or $U_{Eu}$. Furthermore, the Hubbard-$U_{Eu}$ acts only on the 4f shell of the Eu ion, which has a smaller radius than the fully occupied 5s and 5p shells, making a direct effect on the phonon frequencies less likely. 

These observations, along with the fact that magnetism originates from the unpaired electrons on Eu, suggests that the Eu-f states may play a role in the origin of spin-lattice coupling. To probe this further, next we take a closer look at the spin-dependent hybridization of Eu-f electrons with other orbitals and with its effect on the soft-mode behavior.

%=====================================
\begin{figure*}[t]
  \begin{center}
    \includegraphics[width=0.8\hsize]{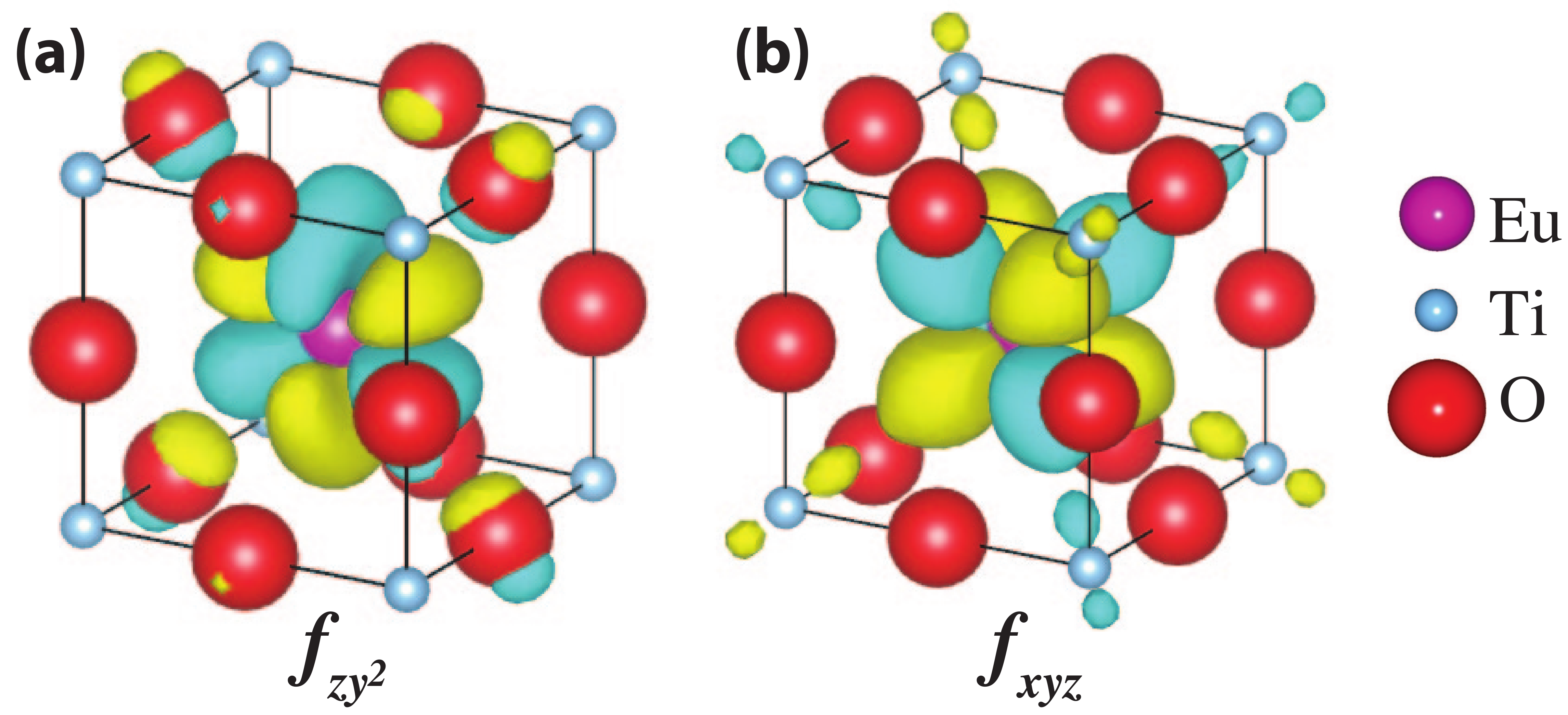}
  \end{center}
    \caption{Two examples of maximally localized Wannier functions of Eu f electrons in EuTiO$_3$. (a) $f_{zy^2}\sim z(4y^2-z^2-x^2)$, (b) $f_{xyz}\sim xyz$. Yellow and green parts of the Wannier Function correspond to isosurfaces of opposite sign, and the Europium ion is in the center of the cubic cell.}
    \label{fig:wannier_example}
\end{figure*}
%=====================================

%=============================
\subsection{Magnetic order control of  Eu-f/Ti-d/O-p hybridization}
\label{hybrid}
%==============================

Despite the small radii of the Eu 4f orbitals, there are no bands purely of Eu character. This is seen in the site-projected density of states (DOS) as shown in Fig.~\ref{fig:dos}a. The DOS peak right below the Fermi level is of dominantly Eu-f character, i.e., the wavefunctions in the energy window corresponding to this peak are mostly localized on the Eu ions, and have the symmetry of f states. There is, however, a small but non-zero contribution from Ti and O ions to this peak as well, indicating that the Eu-f states hybridize with both Ti and O atomic orbitals.\cite{akamatsu2011} This becomes strikingly clear  by considering maximally localized Wannier functions (MLWFs).\cite{WANNIER_REVIEW} In Figure \ref{fig:wannier_example}, we show two examples of occupied Eu f MLWFs, $f_{zy^2}$ and $f_{xyz}$ Wannier orbitals.~\footnote{All of the Wannier functions presented in this study are calculated using an energy range that covers only the Eu f bands. No unentanglement is required since these bands are well separated in energy from others.}
The cubic harmonics corresponding to these orbitals are proportional to $z\left(4y^2-x^2-z^2\right)$ and $xyz$ respectively.

Of particular interest is the  Eu-ion's $f_{xyz}$ Wannier orbital, Fig.~\ref{fig:wannier_example}b, which has lobes directed towards the Ti cation. Notice that the MLWF is mostly localized around the ion's core. 
There is, however, a small, but nonzero, weight around the neighboring Ti ions. This is an explicit  sign that this Eu state hybridizes with a nominally empty Ti-d state(s). This hybridization is important for several reasons, e.g., it has been shown previously that it leads to a superexhange interaction mediated through the Ti cations.~\cite{ranjan2009, akamatsu2011, birol2012} 

\begin{figure}[b]
  \begin{center}
    \includegraphics[width=1.0\hsize]{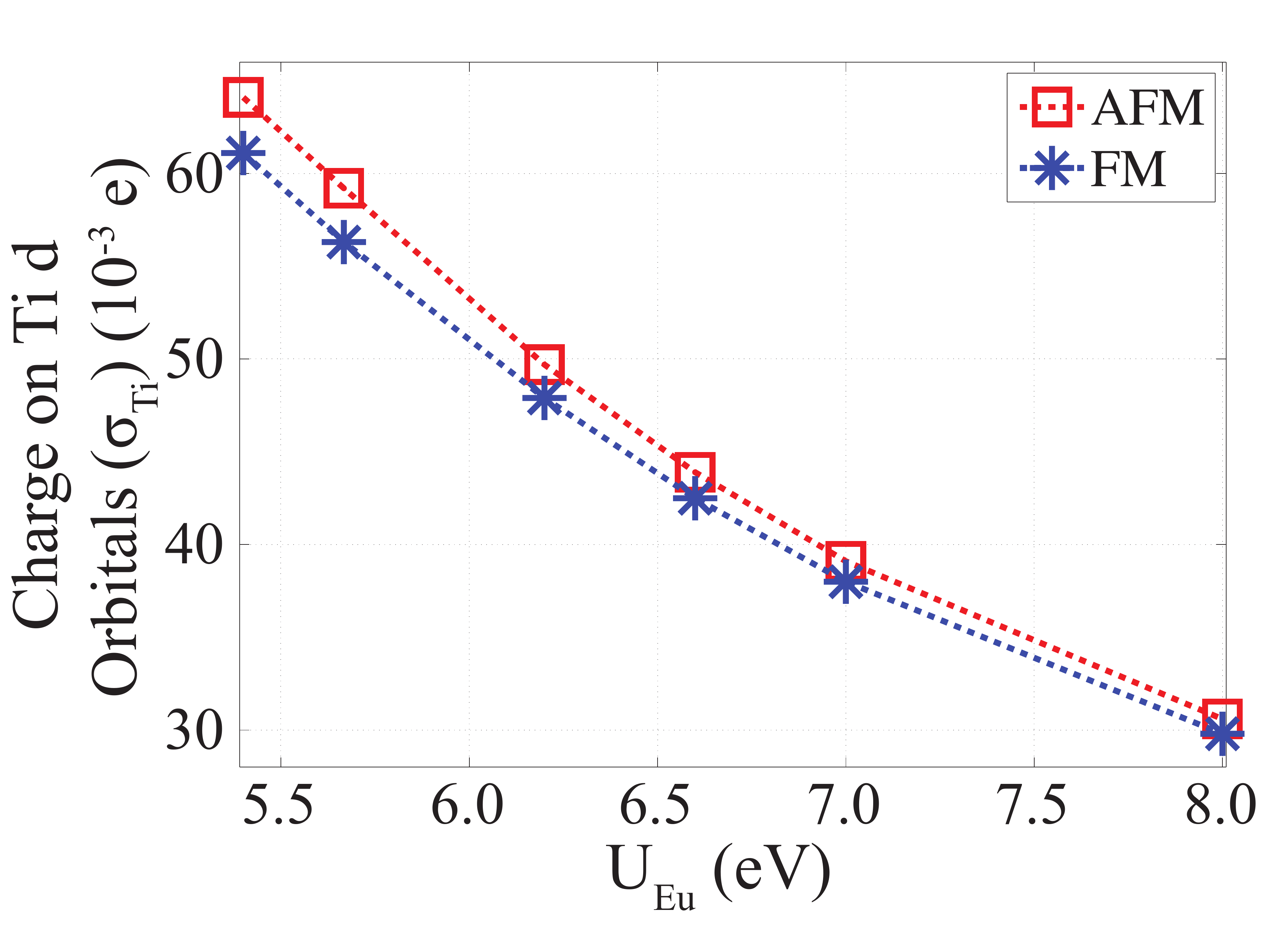}
  \end{center}
    \caption{ Charge on Ti d orbitals due to hybridization of the Eu f states ($\sigma_{Ti}$) versus Hubbard-$U_{Eu}$ used in DFT+U calculations. Red squares and blue asterisks denote values in AFM and FM states respectively. }
    \label{fig:chargeti}
\end{figure}

Our interest here is in the fact that this component of the MLWF  can be thought  as representing the partial occupation of  the Ti-d orbitals. 
It turns out that the dependance of this hybridization on a Hubbard-U applied to the Eu-f states, $U_{Eu}$, brings out the underlying  physics of spin-phonon coupling in EuTiO$_3$. In Fig.~\ref{fig:chargeti}, we plot the total charge within the Ti-d manifold of states due to the hybridization with Eu-f states ($\sigma_{Ti}$) as a function of $U_{Eu}$, considering both the ground state G-type antiferromagnetic and ferromagnetic spin configurations. We obtain $\sigma_{Ti}$ by integrating the DOS projected onto the Ti-d shell over the energy window corresponding to the Eu-f bands (between $\approx-0.5$ and $\approx 0.0$ eV). This  quantity, $\sigma_{Ti}$,   is also related to the weight of the Eu-f MLWFs localized on a Ti site seen in Fig.~\ref{fig:wannier_example}b. Although $\sigma_{Ti}$ is small, there are two clear trends that are evident in Fig.~\ref{fig:chargeti}.: (i) $\sigma_{Ti}$ decreases with increasing $U_{Eu}$ and (ii) $\sigma_{Ti}$ is larger in the AFM state compared to the FM one.\footnote{While the exact quantitative value of $\sigma_{Ti}$ depends on the details of the procedure used to calculate the site projected DOS, such as the radius of the spheres used; the two trends that we report are robust against changes in the sphere size.} 

%=============================
%=============================

%=============================
\begin{figure*}
  \begin{center}
    \includegraphics[width=0.6\hsize]{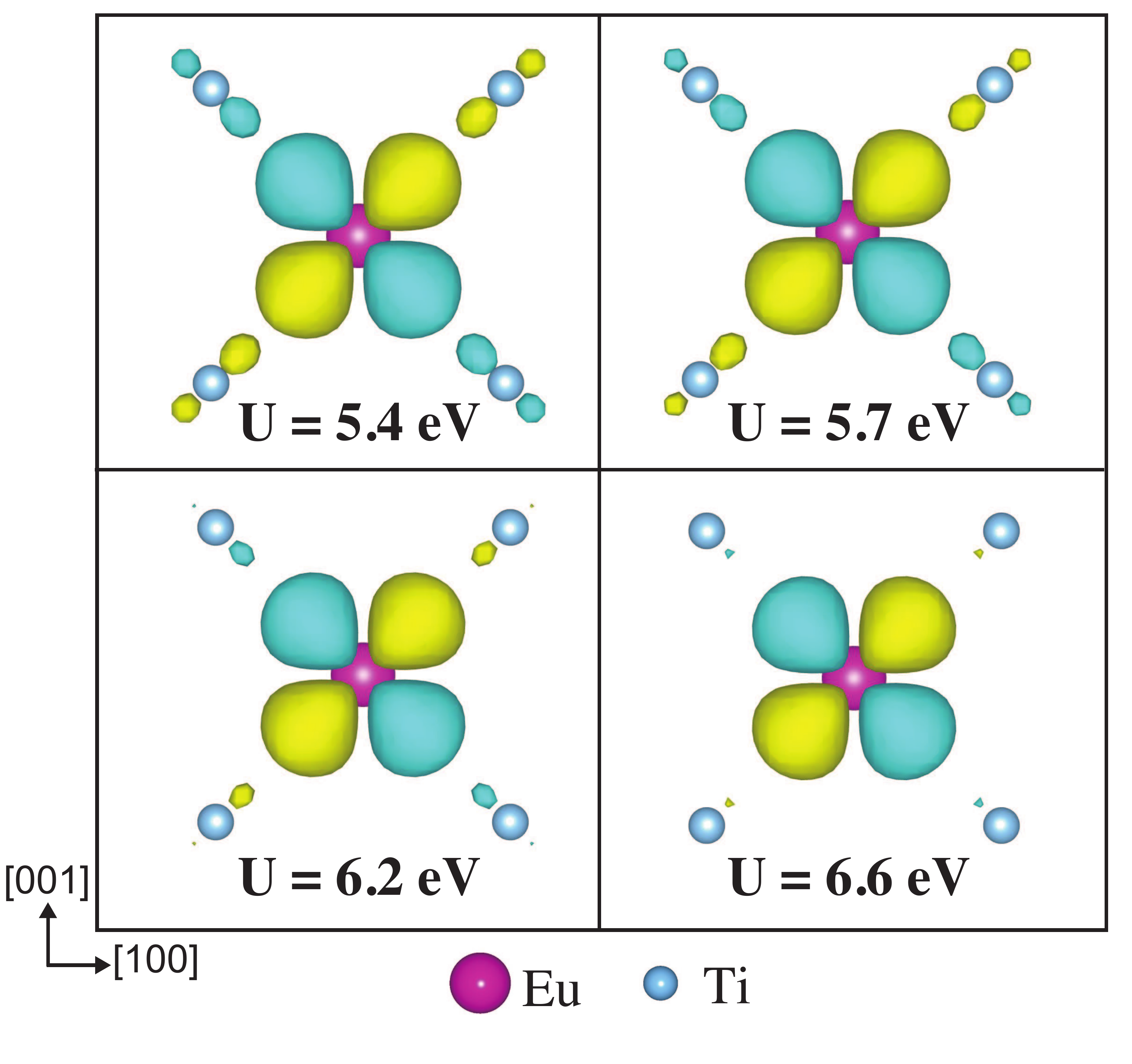}
  \end{center}
    \caption{The Eu $f_{xyz}$ MLWF for different values of $U_{Eu}$. For simplicity, the oxygen ions are not shown on the figure.}
    \label{fig:wannieru}
\end{figure*}
%=============================

The latter will be explained in the proceeding Section while the former, a change in the amount of hybridization with increasing $U_{Eu}$, is not surprising. 
Adding a Coulomb interaction, U, to DFT   causes the corresponding orbitals to become more local. Increasing $U_{Eu}$ makes it energetically favorable for electrons to remain localized in  Eu-f orbitals. 
Also, as observed in Ref. \onlinecite{ranjan2007}, the Eu-f bands move away in energy from the conduction band when $U_{Eu}$ is increased.
As a result,  the hybridization between the Eu-f and the Ti-d states  decreases. This is also clearly visible in Fig.~\ref{fig:wannieru}, where the $f_{xyz}$ MLWF is plotted for different values of $U_{Eu}$. As the Coulomb interaction increases, the lobes localized near the Ti cation get smaller and eventually disappear, consistent with a decreasing $\sigma_{Ti}$.

%=====================================================
\begin{figure}
  \begin{center}
    \includegraphics[width=1\hsize]{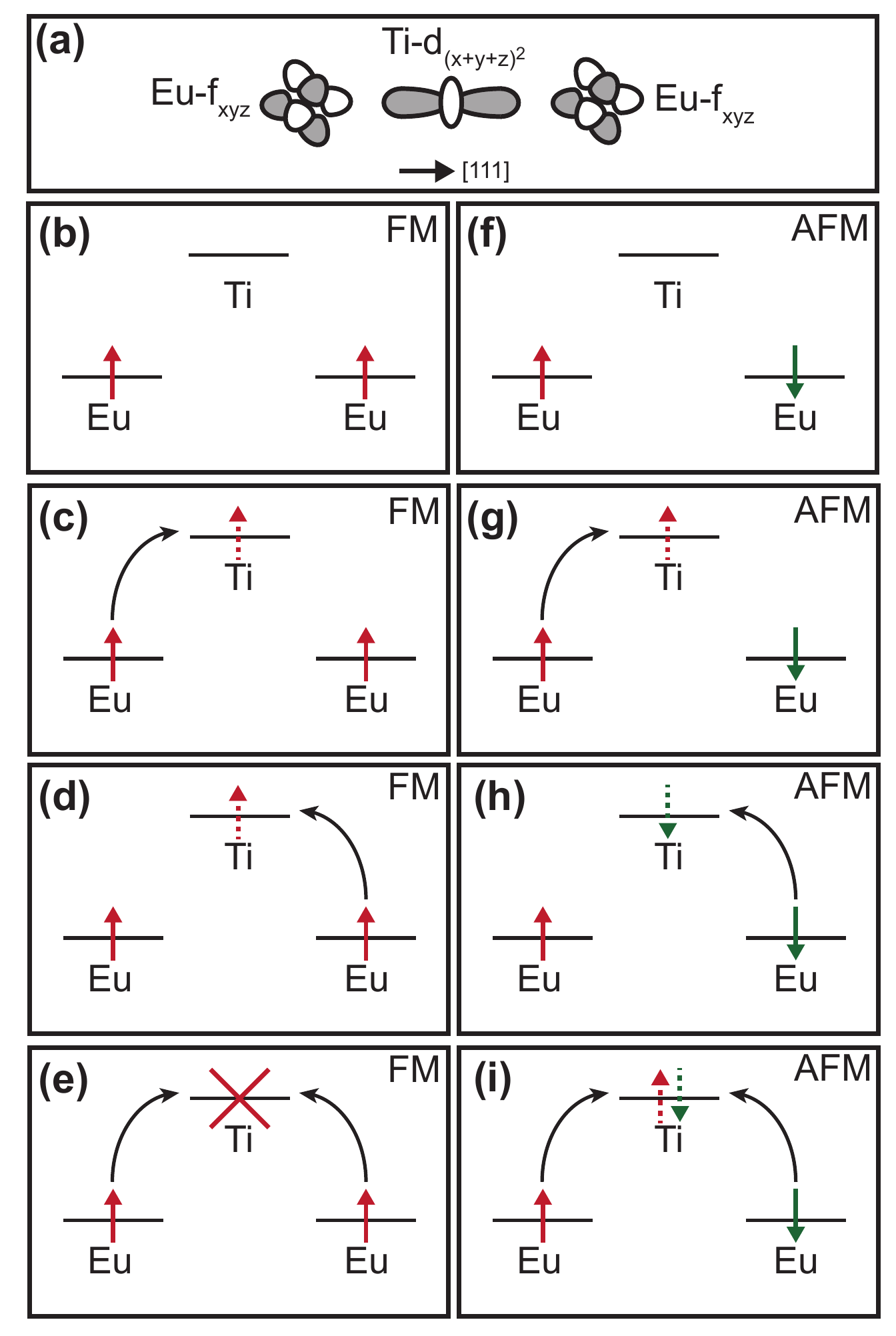}
  \end{center}
    \caption{(a) Sketch of the $f_{xyz}$ orbitals on $3^{rd}$ neighbor Eu ions and the intermediate Ti ion's $d_{(x+y+z)^2}$ orbital. (b) Energy levels of the three orbitals in the FM state. Lowest lying excitations where an electron hops onto the Ti cation, (c) and (d), are allowed, but not both the electrons can hop at the same time because of Pauli exclusion (e). However, in the AFM state, (f), not only the lowest excitations (g) and (h) but also the correlated hopping sketched in (i) is allowed. 
    }
    \label{fig:hybridsketch}
\end{figure}
%=====================================================

%=============================
\subsection{The suppression of ferroelectricity}
\label{ferro}
%==============================
Ferroelectricity in prototypical perovskite ferroelectrics such as BaTiO$_3$ originates from a ``cross-gap'' hybridization~\cite{singh2006} of a  cation's  empty orbitals at the bottom of the conduction band (typically either a transition metal cation's d-orbitals, e.g., Ti$^{4+}$,  or a lone pair active cation's p-orbitals, e.g., Bi$^{3+}$) and the occupied p states of the oxygens at the top of the valence band. This mechanism can be thought of as a second order Jahn-Teller like process.~\cite{bersuker1978, cohen1992,rabe2007, bersuker2012}
This is the mechanism for ferroelectricity in strained EuTiO$_3$. Here, the displacement of a Ti cation towards one of the oxygens increases the Ti-3d/O-2p hybridization, thereby moving the hybridized empty states to higher energies, while lowering the energy of the hybridized occupied states. This ``rehybridization'' leads to a second order energy gain favoring ferroelectricity. If, however, the transition metal d states are partially occupied, there is an extra energy cost associated with moving these states to higher energies, and the tendency towards ferroelectricity is reduced.\cite{khomskii2006} This argument has been mentioned often in the context of the incompatibility of ferroelectricity with B-site magnetism,\cite{hill2000} and is central to both the suppression of ferroelectricity and the origin of spin-lattice coupling in EuTiO$_3$. But first we must understand why $\sigma_{Ti}$ is larger in the AFM state than in a state with parallel spins (FM).

Consider the Eu-Ti-Eu exchange pathway along the [111] direction. In bulk EuTiO$_3$ these Eu cations have a strong AFM interaction, which leads to the observation of (predominantly) G-type magnetic order.\cite{akamatsu2012, akamatsu2013, ryan2013} In Fig.~\ref{fig:hybridsketch}a, the $f_{xyz}$ orbitals on two neighbor spin-polarized Eu$^{2+}$ cations, and the $d_{(x+y+z)^2}$ orbital on the intermediate non-magnetic Ti$^{4+}$ cation are shown. (This particular $d$ orbital has lobes directed towards both  Eu cations and therefore will have the largest hopping to/from the $f_{xyz}$ orbitals.)

First, imagine that the Eu spins were aligned parallel (FM), rather than in the observed AFM configuration. In Fig. \ref{fig:hybridsketch}b, we sketch the energy levels of the three orbitals in this state. Notice that an electron from either Eu cation is allowed by symmetry to hop to the Ti cation (Fig. \ref{fig:hybridsketch}c and d), but that the higher order process where both electrons hop to the Ti atom simultaneously (Fig. \ref{fig:hybridsketch}e) is not allowed due to Pauli exclusion principle. Next consider the same process but with the spins aligned antiparallel (AFM) (Fig. \ref{fig:hybridsketch}f). Now, in addition to the two individual hopping processes, Fig. \ref{fig:hybridsketch}g-h,  the correlated hopping process in Fig.~\ref{fig:hybridsketch}i is allowed, leading to a larger hybridization, thus, a larger $\sigma_{Ti}$ in the AFM state. Combining the physics represented in Figure~\ref{fig:hybridsketch} with that of the rehybridization  mechanism of ferroelectricity leads to a straightforward explanation for the suppression of ferroelectricity and the origin of spin-lattice coupling in EuTiO$_3$.\footnote{Note that DFT with the LDA or GGA approximations is essentially a mean field theory \cite{DFTU:LDAUTYPE1_2} and as a result such correlated processes are not included in it. However, the requirement that the Kohn-Sham states (and the corresponding Wannier states) are orthonormal essentially leads to the same result that if the Eu spins are antiparallel the $f$ electrons can delocalize to the $d$ states of the Ti ion more.} 

As a thought experiment, initially  consider bulk EuTiO$_3$ in which the Eu-f/Ti-d hybridization was removed. One way this can be done from first-principles is by putting the f-electrons in the core of the PAW potential. In this case, the Ti-d states are essentially empty  and are free to hybridize with the O-2p states as the Ti$^{4+}$ cations move off-center, creating a polar lattice distortion is a second-order Jahn-Teller process.  In this case, EuTiO$_3$ should have a ferroelectric instability as in SrTiO$_3$. Our calculations  directly confirm this. Turning on  the Eu-f/Ti-d hybridization, by moving the f-eletroncs from the core of the PAW potential to the valence, increases the occupancy of the Ti d states.
This lowers the energy gained from the 2nd order Jahn-Teller thereby decreasing the tendency towards ferroelectricity, and thus hardening the polar soft-mode. 
This not only explains the suppression of ferroelectricity in EuTiO$_3$ but also our previous result shown in Fig.~\ref{fig:softmode} (another ``dial'' one can turn to  remove, albeit partially, the Eu-f/Ti-d hybridization, and thus increase the tendency towards ferroelectricity, is to increase the Hubbard-U applied to the Eu f states, $U_{Eu}$.)

%=============================
\begin{figure}[b]
  \begin{center}
    \includegraphics[width=1.1\hsize]{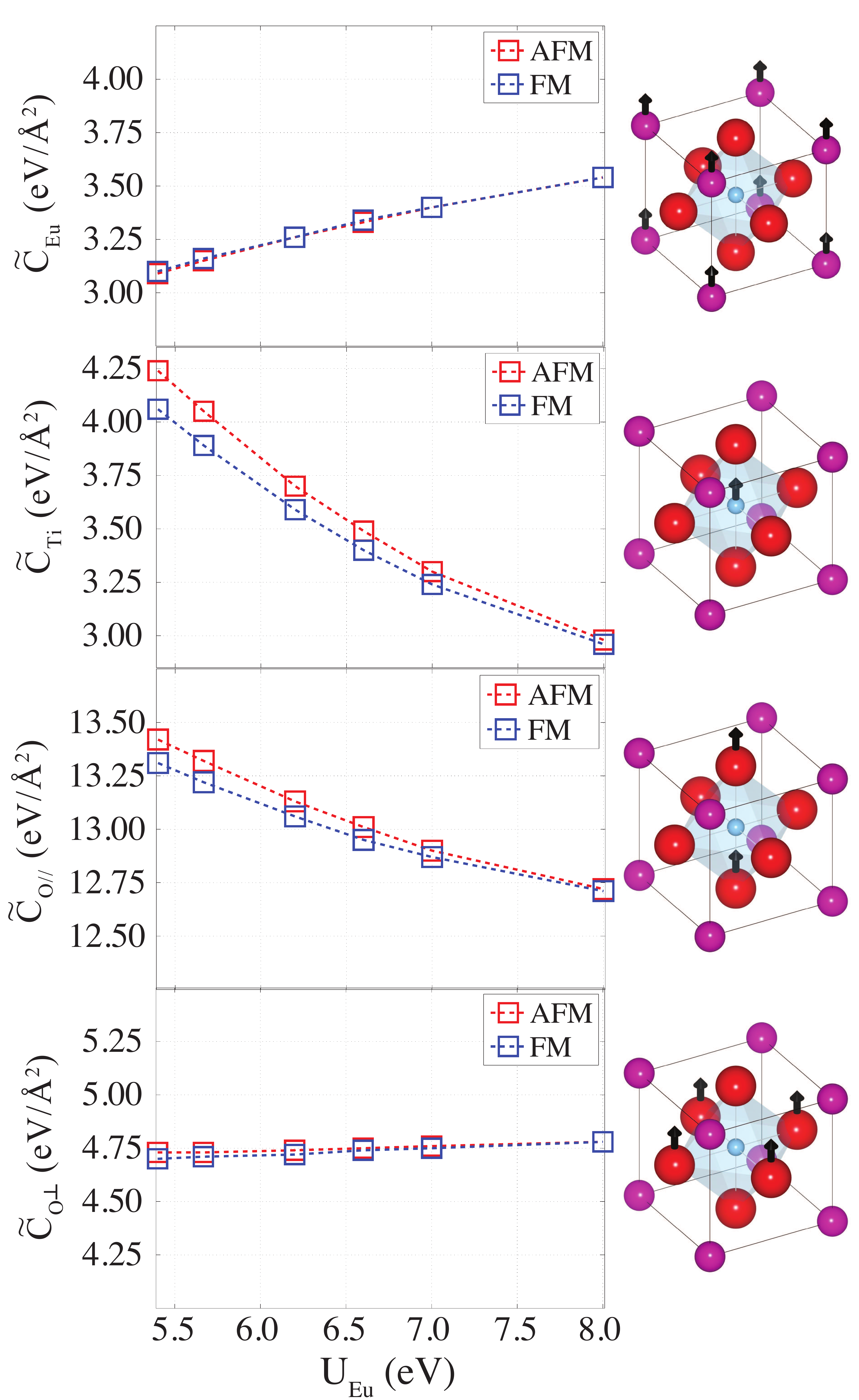}
  \end{center}
    \caption{Self force constants ($\tilde{C}$) for the four $\Gamma$ point symmetry adapted modes (left) and the corresponding displacement patterns (right). Red squares and blue asterisks denote the force constants in AFM and FM states respectively. }
    \label{fig:fcu}
\end{figure}
%=============================

In order to lend support for this scenario, the self force constants, $\tilde{C}$ (i.e., the second derivatives of the total energy with respect to the corresponding ionic displacements) of the 4 symmetry adapted modes of the infrared-active (IR-active) irrep are plotted as a function of $U_{Eu}$ in Fig. \ref{fig:fcu}. 
First note that only  $\tilde{C}_{Ti}$ and $\tilde{C}_{O\parallel}$, which are the only symmetry adapted modes that lead to a first order change in Ti-O distances, decrease significantly with increasing $U_{Eu}$, while $\tilde{C}_{Eu}$ actually increases. Therefore, the softening of $\omega_{SM}$ with increasing $U_{Eu}$ primarily comes  from the softening of the  relative motion of Ti moving against ${O\parallel}$ and not from the Eu motion.
These observations support the claim that the phonon softening with increasing $U_{Eu}$ is a consequence of decreasing Eu-Ti hybridization. 

One question that is natural to ask at this point is whether the emergence of ferroelectricity in EuTiO$_3$ films under biaxial strain\cite{fennie2006, lee2010} is related to a decrease in the Eu-f/Ti-d hybridization. In order to check this possibility, we calculated the DOS and $\sigma_{Ti}$ for EuTiO$_3$ under biaxial strain, not taking into account oxygen octahedral rotations or polarization. The results (not shown) indicate that while $\sigma_{Ti}$ indeed depends on the strain, the change in $\sigma_{Ti}$ for reasonable values of strain is no larger than few percent. 
Thus the emergence of ferroelectricity in EuTiO$_3$ under strain is likely to be of similar nature to that in strained SrTiO$_3$ and CaMnO$_3$.\cite{haeni2004,bhattacharjee2009}

%=============================
\subsection{The mechanism of spin-lattice coupling and the origin of ferromagnetism in strain-induced ferroelectric EuTiO$_3$ }
\label{spin_phonon}
%==============================
%

If the spins in EuTiO$_3$ could be aligned in a parallel direction, e.g., in the presence of a strong  magnetic field, the system would respond by decreasing the Eu-f/Ti-d hybridization, which would subsequently  decrease the occupancy of the Ti-d states,  $\sigma_{Ti}$,   and result in a softening of the polar soft-mode. 
As a result, $\tilde{C}_{Ti}$ and $\tilde{C}_{O\parallel}$ should (and do) have a significant FM-AFM splitting, while $\tilde{C}_{Eu}$ and $\tilde{C}_{O\perp}$ have none, as clearly seen in  Fig. \ref{fig:fcu}. Also note that the splittings of $\tilde{C}_{Ti}$ and $\tilde{C}_{O\parallel}$ decrease with increasing $U_{Eu}$, since $\sigma_{Ti}$ decreases. 
This is the microscopic origin of the spin-phonon observations of Katsufuji and Takagi.
(Note that the AFM-FM splitting of $\omega_{SM}$ does not decrease significantly with  $U_{Eu}$, Fig.~\ref{fig:softmode}, because the eigenvector changes.)

Now imagine that the tendency towards ferroelectricity is greatly increased so as to dominate over the electronic energy gained from the Eu-f/Ti-d hybridization. The system  would respond by decreasing the occupancy of the Ti-d states,  $\sigma_{Ti}$, so as to further increase the energy gain from the polar lattice distortion. This is accomplished by decreasing the Eu-f/Ti-d hybridization, thereby promoting FM interactions between the spins.

\begin{figure}
  \begin{center}
    \includegraphics[width=0.9\hsize]{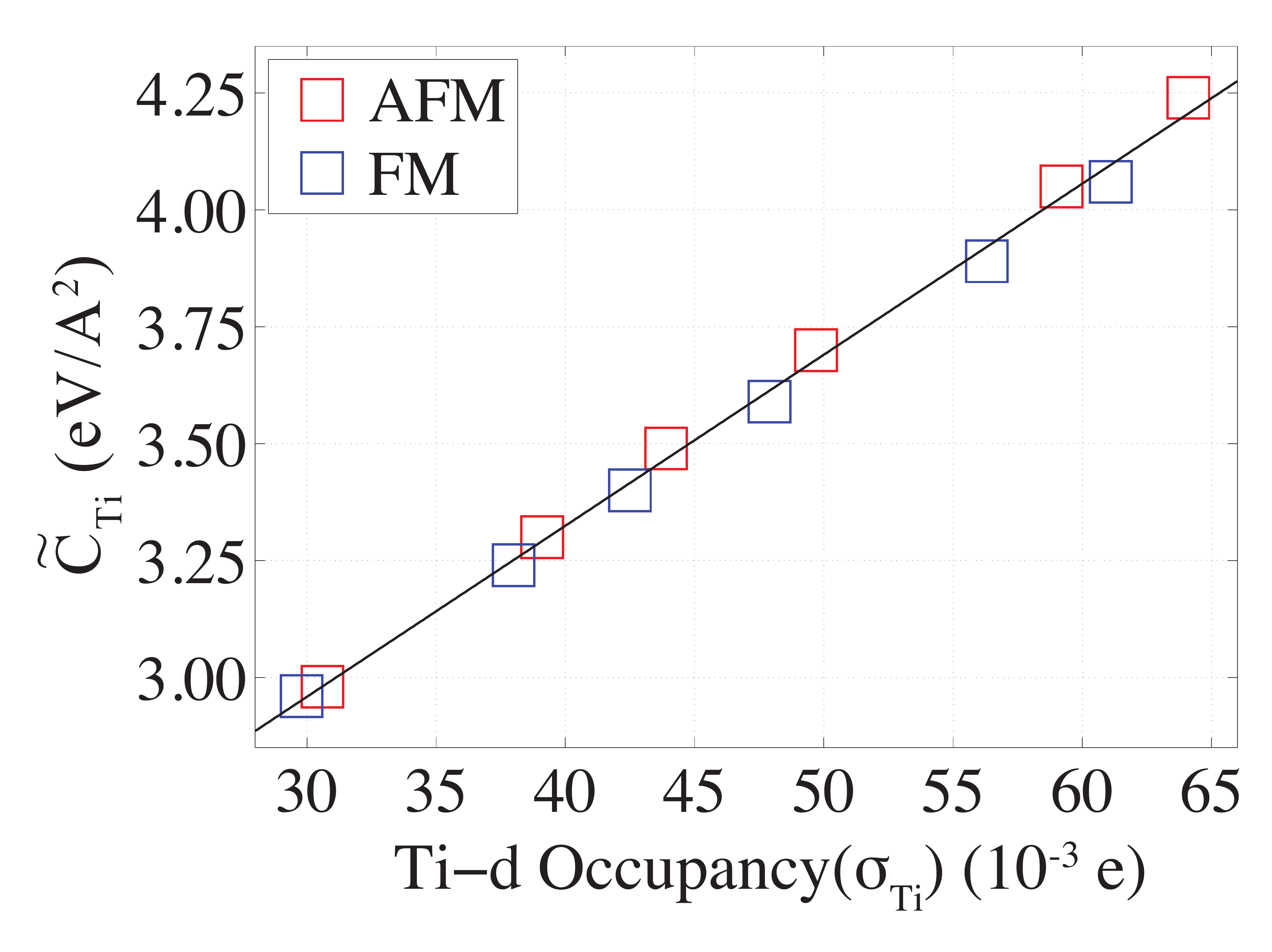}
  \end{center}
    \caption{Self force constant of Ti ion ($\tilde{C}_{Ti}$) as a function of $\sigma_{Ti}$ (the charge on Ti d shell due to hybridization with Eu f states). Red and blue curves denote values calculated in AFM and FM states respectively. Black line is a best fit to the data. Note that the data on this plot can be extracted from Fig. \ref{fig:chargeti} and \ref{fig:fcu}.}
    \label{fig:fcsigma}
\end{figure}

As an additional cross-check, in Fig. \ref{fig:fcsigma} we plot $\tilde{C}_{Ti}$ as a function of $\sigma_{Ti}$. The self force constant of Ti is seen to have an almost linear dependence on $\sigma_{Ti}$ and more importantly, it does not depend on the particular magnetic order, FM or AFM. This universal behavior indicates that the dominant factor determining the change in $\tilde{C}_{Ti}$, and therefore $\omega_{SM}$, is indeed $\sigma_{Ti}$. 
\footnote{Note, however, that the data obtained from the calculations in the AFM state (red) have a slightly larger slope than the one obtained from calculations in the FM state (blue). This indicates that while there are other contributions to spin-phonon coupling apart from the mechanism discussed in this study, they are relatively small.}

As a final check of the validity of our arguments we add a Hubbard-U on the Ti d orbitals as well. Increasing the energy cost of occupying Ti d states suppresses both the spin-phonon coupling and the dependence of $\omega_{SM}$ on $U_{Eu}$ as expected. The strong dependence of spin-phonon coupling to the energy of Ti d states explains why similar spin-lattice physics has not been observed in other compounds similar to EuTiO$_3$, such as EuZrO$_3$.\cite{kolodiazhnyi2010}
(Our calculations for cubic EuZrO$_3$ and EuHfO$_3$ indicate that the spin-phonon coupling in these materials is smaller than the numerical error, in line with our arguments.)

%==============================
\subsection{Oxygen octahedral rotations}
\label{rotations}
%==============================
%
The second question we posed in the introduction concerned the much stronger energy scale associated with  rotations of the octahedra  in EuTiO$_3$ compared with those in SrTiO$_3$. The experimentally measured octahedral rotation angle is also much larger in EuTiO$_3$ (3.6$^\circ$) compared to SrTiO$_3$ (2.1$^\circ$).\cite{sai2000, goian2012, allieta2012}
As we now discuss, this also can be answered by considering the effect of Eu-f states.

Woodward, in his seminal work,\cite{woodward1997} showed that covalent bonding between the A-site cation and the oxygen anions has a stabilizing effect on the oxygen octahedral rotations in perovskites. The octahedral rotations change the coordination environment of the A-site and as a result significantly alter the covalent bonding strength and hybridization between the A-site cation and the oxygens. Akamatsu et al$.$ pointed out another important effect of octahedral rotations in EuMO$_3$ (M=Ti, Zr, Hf) perovskites;\cite{akamatsu2013} they increase the overlap between the Eu-f and B-site d orbitals. This results in an enhanced hybridization between these orbitals, which can also be  seen explicitly in the charge density.\cite{akamatsu2013} Just as the increased A--O covalency stabilizes octahedral rotations, this increased A--B hybridization  also lowers the energy and hence stabilizes the rotational lattice distortion. This explains the stronger rotations observed in EuTiO$_3$ compared to SrTiO$_3$.

\begin{figure}
  \begin{center}
    \includegraphics[width=0.9\hsize]{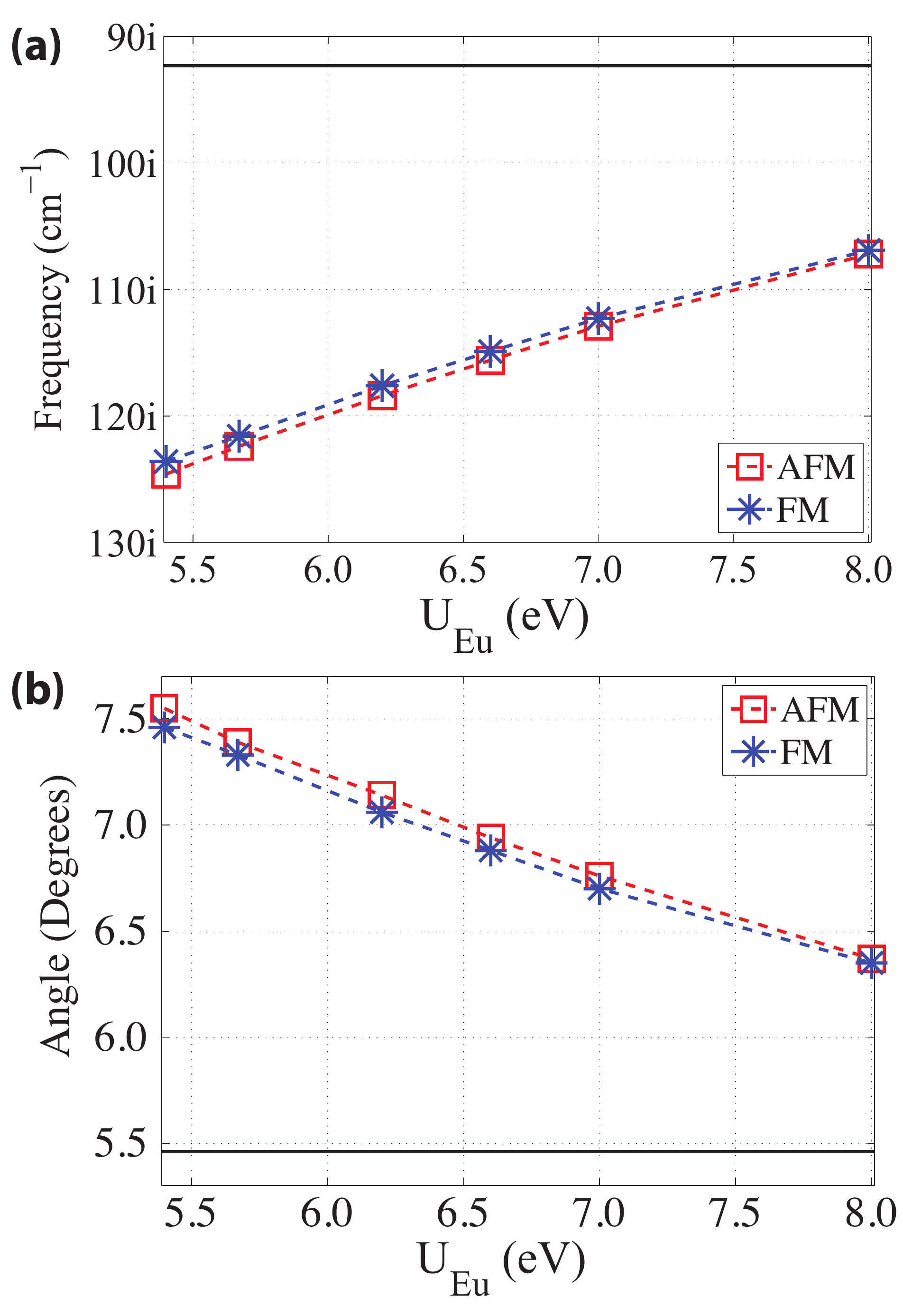}
  \end{center}
    \caption{(a) R point rotation soft mode frequency as a function of $U_{Eu}$ in the FM and the AFM states and the cubic structure with lattice constant fixed to 3.90 \AA. The horizontal black line corresponds to the soft mode frequency in SrTiO$_3$, calculated with the same settings. (b) Octahedral rotation angle, obtained by relaxing the ions in fixed cubic cell, as a function of $U_{Eu}$ in the FM and the AFM states. The horizontal black line corresponds to the rotation angle in SrTiO$_3$, calculated with the same settings. Lines connecting the data points are provided to guide the eye.}
    \label{fig:rotations}
\end{figure}

To help shed  light on this observation, in Fig.~\ref{fig:rotations}a we plot the phonon frequency associated with the R point rotation mode in cubic EuTiO$_3$ as a function of $U_{Eu}$.  With increasing $U_{Eu}$, the f electrons become more localized on the Eu ion, and as a result the stabilizing effect of the Eu-f/Ti-d hybridization decreases. This in turn results in the rotation soft mode becoming more stable (the magnitude of the imaginary frequency decreases). A similar trend is also observed in the ground state octahedral rotation angles reported in Fig.~\ref{fig:rotations}b; the rotation angle decreases with increasing $U_{Eu}$, approaching the value of SrTiO$_3$. In other words, in terms of the octahedral rotations, the behavior of EuTiO$_3$ gets closer to that of SrTiO$_3$ with increasing $U_{Eu}$.\footnote{Note that the  octahedral rotation angles obtained within DFT are roughly 3-4 degrees larger than the experimental value for both EuTiO3 and SrTiO3. This overestimation with respect to experiment within DFT is well-known to occur (see Ref. \onlinecite{sai2000,wahl2008}). Recent high-resolution powder diffraction data shows that in EuTiO$_3$, the local rotation angle is much larger than the average one.\cite{allieta2012} The local value, which is about 8 degrees, agrees well with DFT.\cite{yang2012}}
Note that one should be able to ignore the effect of changing $U_{Eu}$ on the Eu-O covalent bonding as the unoccupied Eu-f states lie at energies much higher than the Fermi level, above the empty Eu-s states. The change in the rotation angle with $U_{Eu}$, therefore, should be attributed solely to the changes in the Eu-f/Ti-d hybridization.

We also performed similar calculations to compare EuZrO$_3$ with SrZrO$_3$, using the same rotation pattern as EuTiO$_3$. The rotation angle difference between these two zirconates is as small as $\approx 0.3^\circ$. This is because EuZrO$_3$ has a much larger band gap than EuTiO$_3$, and as a result there is not a significant hybridization between the Eu-f and the Zr-d states that strengthens the octahedral rotations. The same applies to EuHfO$_3$, which has an octahedral rotation angle of $11.9^\circ$; merely $\sim 0.3^\circ$ degrees larger than SrHfO$_3$. 

%==================================================
\section{Summary}\label{sec:summary}
%==================================================

In summary, by employing DFT+U calculations and MLWFs, we elucidated the key role played by the Eu-f/Ti-d hybridization in EuTiO$_3$. The resultant charge transferred to the Ti-d states ($\sigma_{Ti}$) drives the system away from a ferroelectric (or quantum paraelectric) phase, and causes a dielectric behavior manifestly different from that of SrTiO$_3$. The dependence of $\sigma_{Ti}$ on the magnetic order causes the polar soft-mode frequency to depend on the magnetic state. This is the leading contribution to spin-phonon coupling in EuTiO$_3$. 
Octahedral rotations are also affected from Eu f states' hybridization with Ti d orbitals, and are stronger in EuTiO$_3$ compared to SrTiO$_3$ as a result. 

These results underline the importance of rare earth f electrons in the lattice dynamics and dielectric behavior of TM oxides.\cite{stroppa2010} While the present study is focused on EuTiO$_3$, similar effects could exist in other compounds as well. Taking advantage of the hybridization of rare earth cation with the TM ion might be used as a new knob to tune the system away or close to the ferroelectric transition or perhaps to a  quantum critical point. 

%==================================================
\acknowledgements
%==================================================
We acknowledge fruitful discussions with T.A. Arias, N.A. Benedek, S. Kamba and A. Stroppa. This work is supported by the DOE-BES under Grant No. DE-SCOO02334.


\begin{thebibliography}{79}%
\makeatletter
\providecommand \@ifxundefined [1]{%
 \@ifx{#1\undefined}
}%
\providecommand \@ifnum [1]{%
 \ifnum #1\expandafter \@firstoftwo
 \else \expandafter \@secondoftwo
 \fi
}%
\providecommand \@ifx [1]{%
 \ifx #1\expandafter \@firstoftwo
 \else \expandafter \@secondoftwo
 \fi
}%
\providecommand \natexlab [1]{#1}%
\providecommand \enquote  [1]{``#1''}%
\providecommand \bibnamefont  [1]{#1}%
\providecommand \bibfnamefont [1]{#1}%
\providecommand \citenamefont [1]{#1}%
\providecommand \href@noop [0]{\@secondoftwo}%
\providecommand \href [0]{\begingroup \@sanitize@url \@href}%
\providecommand \@href[1]{\@@startlink{#1}\@@href}%
\providecommand \@@href[1]{\endgroup#1\@@endlink}%
\providecommand \@sanitize@url [0]{\catcode `\\12\catcode `\$12\catcode
  `\&12\catcode `\#12\catcode `\^12\catcode `\_12\catcode `\%12\relax}%
\providecommand \@@startlink[1]{}%
\providecommand \@@endlink[0]{}%
\providecommand \url  [0]{\begingroup\@sanitize@url \@url }%
\providecommand \@url [1]{\endgroup\@href {#1}{\urlprefix }}%
\providecommand \urlprefix  [0]{URL }%
\providecommand \Eprint [0]{\href }%
\providecommand \doibase [0]{http://dx.doi.org/}%
\providecommand \selectlanguage [0]{\@gobble}%
\providecommand \bibinfo  [0]{\@secondoftwo}%
\providecommand \bibfield  [0]{\@secondoftwo}%
\providecommand \translation [1]{[#1]}%
\providecommand \BibitemOpen [0]{}%
\providecommand \bibitemStop [0]{}%
\providecommand \bibitemNoStop [0]{.\EOS\space}%
\providecommand \EOS [0]{\spacefactor3000\relax}%
\providecommand \BibitemShut  [1]{\csname bibitem#1\endcsname}%
\let\auto@bib@innerbib\@empty
%</preamble>
\bibitem [{\citenamefont {Nan}\ \emph {et~al.}(2008)\citenamefont {Nan},
  \citenamefont {Bichurin}, \citenamefont {Dong}, \citenamefont {Viehland},\
  and\ \citenamefont {Srinivasan}}]{nan2008}%
  \BibitemOpen
  \bibfield  {author} {\bibinfo {author} {\bibfnamefont {C.-W.}\ \bibnamefont
  {Nan}}, \bibinfo {author} {\bibfnamefont {M.~I.}\ \bibnamefont {Bichurin}},
  \bibinfo {author} {\bibfnamefont {S.}~\bibnamefont {Dong}}, \bibinfo {author}
  {\bibfnamefont {D.}~\bibnamefont {Viehland}}, \ and\ \bibinfo {author}
  {\bibfnamefont {G.}~\bibnamefont {Srinivasan}},\ }\href@noop {} {\bibfield
  {journal} {\bibinfo  {journal} {Journal of Applied Physics}\ }\textbf
  {\bibinfo {volume} {103}},\ \bibinfo {pages} {031101} (\bibinfo {year}
  {2008})}\BibitemShut {NoStop}%
\bibitem [{\citenamefont {Martin}\ and\ \citenamefont
  {Schlom}(2012)}]{martin2012}%
  \BibitemOpen
  \bibfield  {author} {\bibinfo {author} {\bibfnamefont {L.~W.}\ \bibnamefont
  {Martin}}\ and\ \bibinfo {author} {\bibfnamefont {D.~G.}\ \bibnamefont
  {Schlom}},\ }\href@noop {} {\bibfield  {journal} {\bibinfo  {journal}
  {Current Opinion in Solid State and Materials Science}\ }\textbf {\bibinfo
  {volume} {16}},\ \bibinfo {pages} {199 } (\bibinfo {year}
  {2012})}\BibitemShut {NoStop}%
\bibitem [{\citenamefont {Martin}\ and\ \citenamefont
  {Ramesh}(2012)}]{martin2012b}%
  \BibitemOpen
  \bibfield  {author} {\bibinfo {author} {\bibfnamefont {L.~W.}\ \bibnamefont
  {Martin}}\ and\ \bibinfo {author} {\bibfnamefont {R.}~\bibnamefont
  {Ramesh}},\ }\href@noop {} {\bibfield  {journal} {\bibinfo  {journal} {Acta
  Materialia}\ }\textbf {\bibinfo {volume} {60}},\ \bibinfo {pages} {2449}
  (\bibinfo {year} {2012})}\BibitemShut {NoStop}%
\bibitem [{\citenamefont {He}\ \emph {et~al.}(2012)\citenamefont {He},
  \citenamefont {Arenholz}, \citenamefont {Scholl}, \citenamefont {Chu},\ and\
  \citenamefont {Ramesh}}]{he2012}%
  \BibitemOpen
  \bibfield  {author} {\bibinfo {author} {\bibfnamefont {Q.}~\bibnamefont
  {He}}, \bibinfo {author} {\bibfnamefont {E.}~\bibnamefont {Arenholz}},
  \bibinfo {author} {\bibfnamefont {A.}~\bibnamefont {Scholl}}, \bibinfo
  {author} {\bibfnamefont {Y.-H.}\ \bibnamefont {Chu}}, \ and\ \bibinfo
  {author} {\bibfnamefont {R.}~\bibnamefont {Ramesh}},\ }\href@noop {}
  {\bibfield  {journal} {\bibinfo  {journal} {Current Opinion in Solid State
  and Materials Science}\ }\textbf {\bibinfo {volume} {16}},\ \bibinfo {pages}
  {216 } (\bibinfo {year} {2012})}\BibitemShut {NoStop}%
\bibitem [{\citenamefont {Cohen}(1993)}]{cohen1993}%
  \BibitemOpen
  \bibfield  {author} {\bibinfo {author} {\bibfnamefont {M.}~\bibnamefont
  {Cohen}},\ }\href@noop {} {\bibfield  {journal} {\bibinfo  {journal}
  {Science}\ }\textbf {\bibinfo {volume} {261}},\ \bibinfo {pages} {307}
  (\bibinfo {year} {1993})}\BibitemShut {NoStop}%
\bibitem [{\citenamefont {Franceschetti}\ and\ \citenamefont
  {Zunger}(1999)}]{franceschetti1999}%
  \BibitemOpen
  \bibfield  {author} {\bibinfo {author} {\bibfnamefont {A.}~\bibnamefont
  {Franceschetti}}\ and\ \bibinfo {author} {\bibfnamefont {A.}~\bibnamefont
  {Zunger}},\ }\href@noop {} {\bibfield  {journal} {\bibinfo  {journal}
  {Nature}\ }\textbf {\bibinfo {volume} {402}},\ \bibinfo {pages} {60}
  (\bibinfo {year} {1999})}\BibitemShut {NoStop}%
\bibitem [{\citenamefont {Spaldin}\ and\ \citenamefont
  {Pickett}(2003)}]{spaldin2003}%
  \BibitemOpen
  \bibfield  {author} {\bibinfo {author} {\bibfnamefont {N.~A.}\ \bibnamefont
  {Spaldin}}\ and\ \bibinfo {author} {\bibfnamefont {W.~E.}\ \bibnamefont
  {Pickett}},\ }\href@noop {} {\bibfield  {journal} {\bibinfo  {journal}
  {Journal of Solid State Chemistry}\ }\textbf {\bibinfo {volume} {176}},\
  \bibinfo {pages} {615 } (\bibinfo {year} {2003})}\BibitemShut {NoStop}%
\bibitem [{\citenamefont {Hafner}\ \emph {et~al.}(2006)\citenamefont {Hafner},
  \citenamefont {Wolverton},\ and\ \citenamefont {Ceder}}]{hafner2006}%
  \BibitemOpen
  \bibfield  {author} {\bibinfo {author} {\bibfnamefont {J.}~\bibnamefont
  {Hafner}}, \bibinfo {author} {\bibfnamefont {C.}~\bibnamefont {Wolverton}}, \
  and\ \bibinfo {author} {\bibfnamefont {G.}~\bibnamefont {Ceder}},\
  }\href@noop {} {\bibfield  {journal} {\bibinfo  {journal} {MRS Bulletin}\
  }\textbf {\bibinfo {volume} {31}},\ \bibinfo {pages} {659} (\bibinfo {year}
  {2006})}\BibitemShut {NoStop}%
\bibitem [{\citenamefont {Benedek}\ and\ \citenamefont
  {Fennie}(2011)}]{benedek2011}%
  \BibitemOpen
  \bibfield  {author} {\bibinfo {author} {\bibfnamefont {N.~A.}\ \bibnamefont
  {Benedek}}\ and\ \bibinfo {author} {\bibfnamefont {C.~J.}\ \bibnamefont
  {Fennie}},\ }\href@noop {} {\bibfield  {journal} {\bibinfo  {journal} {Phys.
  Rev. Lett.}\ }\textbf {\bibinfo {volume} {106}},\ \bibinfo {pages} {107204}
  (\bibinfo {year} {2011})}\BibitemShut {NoStop}%
\bibitem [{\citenamefont {Bousquet}\ and\ \citenamefont
  {Spaldin}(2011)}]{bousquet2011}%
  \BibitemOpen
  \bibfield  {author} {\bibinfo {author} {\bibfnamefont {E.}~\bibnamefont
  {Bousquet}}\ and\ \bibinfo {author} {\bibfnamefont {N.}~\bibnamefont
  {Spaldin}},\ }\href@noop {} {\bibfield  {journal} {\bibinfo  {journal} {Phys.
  Rev. Lett.}\ }\textbf {\bibinfo {volume} {107}},\ \bibinfo {pages} {197603}
  (\bibinfo {year} {2011})}\BibitemShut {NoStop}%
\bibitem [{\citenamefont {Birol}\ \emph {et~al.}(2012)\citenamefont {Birol},
  \citenamefont {Benedek}, \citenamefont {Das}, \citenamefont {Wysocki},
  \citenamefont {Mulder}, \citenamefont {Abbett}, \citenamefont {Smith},
  \citenamefont {Ghosh},\ and\ \citenamefont {Fennie}}]{birol2012}%
  \BibitemOpen
  \bibfield  {author} {\bibinfo {author} {\bibfnamefont {T.}~\bibnamefont
  {Birol}}, \bibinfo {author} {\bibfnamefont {N.~A.}\ \bibnamefont {Benedek}},
  \bibinfo {author} {\bibfnamefont {H.}~\bibnamefont {Das}}, \bibinfo {author}
  {\bibfnamefont {A.~L.}\ \bibnamefont {Wysocki}}, \bibinfo {author}
  {\bibfnamefont {A.~T.}\ \bibnamefont {Mulder}}, \bibinfo {author}
  {\bibfnamefont {B.~M.}\ \bibnamefont {Abbett}}, \bibinfo {author}
  {\bibfnamefont {E.~H.}\ \bibnamefont {Smith}}, \bibinfo {author}
  {\bibfnamefont {S.}~\bibnamefont {Ghosh}}, \ and\ \bibinfo {author}
  {\bibfnamefont {C.~J.}\ \bibnamefont {Fennie}},\ }\href@noop {} {\bibfield
  {journal} {\bibinfo  {journal} {Current Opinion in Solid State and Materials
  Science}\ }\textbf {\bibinfo {volume} {16}},\ \bibinfo {pages} {227 }
  (\bibinfo {year} {2012})}\BibitemShut {NoStop}%
\bibitem [{\citenamefont {Picozzi}\ and\ \citenamefont
  {Stroppa}(2012)}]{picozzi2012}%
  \BibitemOpen
  \bibfield  {author} {\bibinfo {author} {\bibfnamefont {S.}~\bibnamefont
  {Picozzi}}\ and\ \bibinfo {author} {\bibfnamefont {A.}~\bibnamefont
  {Stroppa}},\ }\href@noop {} {\bibfield  {journal} {\bibinfo  {journal}
  {European Physical Journal B}\ }\textbf {\bibinfo {volume} {85}},\ \bibinfo
  {pages} {1} (\bibinfo {year} {2012})}\BibitemShut {NoStop}%
\bibitem [{\citenamefont {Yin}\ and\ \citenamefont {Kotliar}(2013)}]{yin2013}%
  \BibitemOpen
  \bibfield  {author} {\bibinfo {author} {\bibfnamefont {Z.~P.}\ \bibnamefont
  {Yin}}\ and\ \bibinfo {author} {\bibfnamefont {G.}~\bibnamefont {Kotliar}},\
  }\href@noop {} {\bibfield  {journal} {\bibinfo  {journal} {Europhysics
  Letters}\ }\textbf {\bibinfo {volume} {101}} (\bibinfo {year}
  {2013})}\BibitemShut {NoStop}%
\bibitem [{\citenamefont {Lee}\ and\ \citenamefont {Rabe}(2010)}]{lee2010b}%
  \BibitemOpen
  \bibfield  {author} {\bibinfo {author} {\bibfnamefont {J.~H.}\ \bibnamefont
  {Lee}}\ and\ \bibinfo {author} {\bibfnamefont {K.~M.}\ \bibnamefont {Rabe}},\
  }\href@noop {} {\bibfield  {journal} {\bibinfo  {journal} {Phys. Rev. Lett.}\
  }\textbf {\bibinfo {volume} {104}},\ \bibinfo {pages} {207204} (\bibinfo
  {year} {2010})}\BibitemShut {NoStop}%
\bibitem [{\citenamefont {Lee}\ and\ \citenamefont
  {Rabe}(2011{\natexlab{a}})}]{lee2011b}%
  \BibitemOpen
  \bibfield  {author} {\bibinfo {author} {\bibfnamefont {J.~H.}\ \bibnamefont
  {Lee}}\ and\ \bibinfo {author} {\bibfnamefont {K.~M.}\ \bibnamefont {Rabe}},\
  }\href@noop {} {\bibfield  {journal} {\bibinfo  {journal} {Phys. Rev. Lett.}\
  }\textbf {\bibinfo {volume} {107}},\ \bibinfo {pages} {067601} (\bibinfo
  {year} {2011}{\natexlab{a}})}\BibitemShut {NoStop}%
\bibitem [{\citenamefont {Giovannetti}\ \emph {et~al.}(2012)\citenamefont
  {Giovannetti}, \citenamefont {Kumar}, \citenamefont {Ortix}, \citenamefont
  {Capone},\ and\ \citenamefont {van~den Brink}}]{giovannetti2012}%
  \BibitemOpen
  \bibfield  {author} {\bibinfo {author} {\bibfnamefont {G.}~\bibnamefont
  {Giovannetti}}, \bibinfo {author} {\bibfnamefont {S.}~\bibnamefont {Kumar}},
  \bibinfo {author} {\bibfnamefont {C.}~\bibnamefont {Ortix}}, \bibinfo
  {author} {\bibfnamefont {M.}~\bibnamefont {Capone}}, \ and\ \bibinfo {author}
  {\bibfnamefont {J.}~\bibnamefont {van~den Brink}},\ }\href@noop {} {\bibfield
   {journal} {\bibinfo  {journal} {Phys. Rev. Lett.}\ }\textbf {\bibinfo
  {volume} {109}},\ \bibinfo {pages} {107601} (\bibinfo {year}
  {2012})}\BibitemShut {NoStop}%
\bibitem [{\citenamefont {McGuire}\ \emph {et~al.}(1966)\citenamefont
  {McGuire}, \citenamefont {Shafer}, \citenamefont {Joenk}, \citenamefont
  {Alperin},\ and\ \citenamefont {Pickart}}]{mcguire1966}%
  \BibitemOpen
  \bibfield  {author} {\bibinfo {author} {\bibfnamefont {T.}~\bibnamefont
  {McGuire}}, \bibinfo {author} {\bibfnamefont {M.}~\bibnamefont {Shafer}},
  \bibinfo {author} {\bibfnamefont {R.}~\bibnamefont {Joenk}}, \bibinfo
  {author} {\bibfnamefont {H.}~\bibnamefont {Alperin}}, \ and\ \bibinfo
  {author} {\bibfnamefont {S.}~\bibnamefont {Pickart}},\ }\href@noop {}
  {\bibfield  {journal} {\bibinfo  {journal} {Journal of Applied Physics}\
  }\textbf {\bibinfo {volume} {37}},\ \bibinfo {pages} {981} (\bibinfo {year}
  {1966})}\BibitemShut {NoStop}%
\bibitem [{\citenamefont {Chien}\ \emph {et~al.}(1974)\citenamefont {Chien},
  \citenamefont {DeBenedetti},\ and\ \citenamefont {Barros}}]{chien1974}%
  \BibitemOpen
  \bibfield  {author} {\bibinfo {author} {\bibfnamefont {C.-L.}\ \bibnamefont
  {Chien}}, \bibinfo {author} {\bibfnamefont {S.}~\bibnamefont {DeBenedetti}},
  \ and\ \bibinfo {author} {\bibfnamefont {F.~D.~S.}\ \bibnamefont {Barros}},\
  }\href@noop {} {\bibfield  {journal} {\bibinfo  {journal} {Physical Review
  B}\ }\textbf {\bibinfo {volume} {10}},\ \bibinfo {pages} {3913} (\bibinfo
  {year} {1974})}\BibitemShut {NoStop}%
\bibitem [{\citenamefont {Katsufuji}\ and\ \citenamefont
  {Takagi}(2001)}]{katsufuji2001}%
  \BibitemOpen
  \bibfield  {author} {\bibinfo {author} {\bibfnamefont {T.}~\bibnamefont
  {Katsufuji}}\ and\ \bibinfo {author} {\bibfnamefont {H.}~\bibnamefont
  {Takagi}},\ }\href@noop {} {\bibfield  {journal} {\bibinfo  {journal}
  {Physical Review B}\ }\textbf {\bibinfo {volume} {64}},\ \bibinfo {pages}
  {054415} (\bibinfo {year} {2001})}\BibitemShut {NoStop}%
\bibitem [{\citenamefont {Lee}\ and\ \citenamefont
  {Rabe}(2011{\natexlab{b}})}]{lee2011a}%
  \BibitemOpen
  \bibfield  {author} {\bibinfo {author} {\bibfnamefont {J.~H.}\ \bibnamefont
  {Lee}}\ and\ \bibinfo {author} {\bibfnamefont {K.~M.}\ \bibnamefont {Rabe}},\
  }\href@noop {} {\bibfield  {journal} {\bibinfo  {journal} {Phys. Rev. B}\
  }\textbf {\bibinfo {volume} {84}},\ \bibinfo {pages} {104440} (\bibinfo
  {year} {2011}{\natexlab{b}})}\BibitemShut {NoStop}%
\bibitem [{\citenamefont {Hong}\ \emph {et~al.}(2012)\citenamefont {Hong},
  \citenamefont {Stroppa}, \citenamefont {{\'I}{\~n}iguez}, \citenamefont
  {Picozzi},\ and\ \citenamefont {Vanderbilt}}]{hong2012}%
  \BibitemOpen
  \bibfield  {author} {\bibinfo {author} {\bibfnamefont {J.}~\bibnamefont
  {Hong}}, \bibinfo {author} {\bibfnamefont {A.}~\bibnamefont {Stroppa}},
  \bibinfo {author} {\bibfnamefont {J.}~\bibnamefont {{\'I}{\~n}iguez}},
  \bibinfo {author} {\bibfnamefont {S.}~\bibnamefont {Picozzi}}, \ and\
  \bibinfo {author} {\bibfnamefont {D.}~\bibnamefont {Vanderbilt}},\
  }\href@noop {} {\bibfield  {journal} {\bibinfo  {journal} {Physical Review
  B}\ }\textbf {\bibinfo {volume} {85}},\ \bibinfo {pages} {054417} (\bibinfo
  {year} {2012})}\BibitemShut {NoStop}%
\bibitem [{\citenamefont {Fennie}\ and\ \citenamefont
  {Rabe}(2006)}]{fennie2006}%
  \BibitemOpen
  \bibfield  {author} {\bibinfo {author} {\bibfnamefont {C.~J.}\ \bibnamefont
  {Fennie}}\ and\ \bibinfo {author} {\bibfnamefont {K.~M.}\ \bibnamefont
  {Rabe}},\ }\href@noop {} {\bibfield  {journal} {\bibinfo  {journal} {Phys.
  Rev. Lett.}\ }\textbf {\bibinfo {volume} {97}},\ \bibinfo {pages} {267602}
  (\bibinfo {year} {2006})}\BibitemShut {NoStop}%
\bibitem [{\citenamefont {Ranjan}\ \emph {et~al.}(2007)\citenamefont {Ranjan},
  \citenamefont {Nabi},\ and\ \citenamefont {Pentcheva}}]{ranjan2007}%
  \BibitemOpen
  \bibfield  {author} {\bibinfo {author} {\bibfnamefont {R.}~\bibnamefont
  {Ranjan}}, \bibinfo {author} {\bibfnamefont {H.~S.}\ \bibnamefont {Nabi}}, \
  and\ \bibinfo {author} {\bibfnamefont {R.}~\bibnamefont {Pentcheva}},\
  }\href@noop {} {\bibfield  {journal} {\bibinfo  {journal} {Journal of
  Physics: Condensed Matter}\ }\textbf {\bibinfo {volume} {19}},\ \bibinfo
  {pages} {406217} (\bibinfo {year} {2007})}\BibitemShut {NoStop}%
\bibitem [{\citenamefont {Kamba}\ \emph {et~al.}(2007)\citenamefont {Kamba},
  \citenamefont {Nuzhnyy}, \citenamefont {Van{\v{e}}k}, \citenamefont
  {Savinov}, \citenamefont {Kn{\'\i}{\v{z}}ek}, \citenamefont {Shen},
  \citenamefont {{\v{S}}antav{\'a}}, \citenamefont {Maca}, \citenamefont
  {Sadowski},\ and\ \citenamefont {Petzelt}}]{kamba2007}%
  \BibitemOpen
  \bibfield  {author} {\bibinfo {author} {\bibfnamefont {S.}~\bibnamefont
  {Kamba}}, \bibinfo {author} {\bibfnamefont {D.}~\bibnamefont {Nuzhnyy}},
  \bibinfo {author} {\bibfnamefont {P.}~\bibnamefont {Van{\v{e}}k}}, \bibinfo
  {author} {\bibfnamefont {M.}~\bibnamefont {Savinov}}, \bibinfo {author}
  {\bibfnamefont {K.}~\bibnamefont {Kn{\'\i}{\v{z}}ek}}, \bibinfo {author}
  {\bibfnamefont {Z.}~\bibnamefont {Shen}}, \bibinfo {author} {\bibfnamefont
  {E.}~\bibnamefont {{\v{S}}antav{\'a}}}, \bibinfo {author} {\bibfnamefont
  {K.}~\bibnamefont {Maca}}, \bibinfo {author} {\bibfnamefont {M.}~\bibnamefont
  {Sadowski}}, \ and\ \bibinfo {author} {\bibfnamefont {J.}~\bibnamefont
  {Petzelt}},\ }\href@noop {} {\bibfield  {journal} {\bibinfo  {journal}
  {Europhysics Letters}\ }\textbf {\bibinfo {volume} {80}},\ \bibinfo {pages}
  {27002} (\bibinfo {year} {2007})}\BibitemShut {NoStop}%
\bibitem [{\citenamefont {Kamba}\ \emph {et~al.}(2012)\citenamefont {Kamba},
  \citenamefont {Goian}, \citenamefont {Orlita}, \citenamefont {Nuzhnyy},
  \citenamefont {Lee}, \citenamefont {Schlom}, \citenamefont {Rushchanskii},
  \citenamefont {Le{\v{z}}ai{\'c}}, \citenamefont {Birol}, \citenamefont
  {Fennie} \emph {et~al.}}]{kamba2012a}%
  \BibitemOpen
  \bibfield  {author} {\bibinfo {author} {\bibfnamefont {S.}~\bibnamefont
  {Kamba}}, \bibinfo {author} {\bibfnamefont {V.}~\bibnamefont {Goian}},
  \bibinfo {author} {\bibfnamefont {M.}~\bibnamefont {Orlita}}, \bibinfo
  {author} {\bibfnamefont {D.}~\bibnamefont {Nuzhnyy}}, \bibinfo {author}
  {\bibfnamefont {J.}~\bibnamefont {Lee}}, \bibinfo {author} {\bibfnamefont
  {D.}~\bibnamefont {Schlom}}, \bibinfo {author} {\bibfnamefont
  {K.}~\bibnamefont {Rushchanskii}}, \bibinfo {author} {\bibfnamefont
  {M.}~\bibnamefont {Le{\v{z}}ai{\'c}}}, \bibinfo {author} {\bibfnamefont
  {T.}~\bibnamefont {Birol}}, \bibinfo {author} {\bibfnamefont
  {C.}~\bibnamefont {Fennie}},  \emph {et~al.},\ }\href@noop {} {\bibfield
  {journal} {\bibinfo  {journal} {Physical Review B}\ }\textbf {\bibinfo
  {volume} {85}},\ \bibinfo {pages} {094435} (\bibinfo {year}
  {2012})}\BibitemShut {NoStop}%
\bibitem [{\citenamefont {Newnham}(1998)}]{newnham1998}%
  \BibitemOpen
  \bibfield  {author} {\bibinfo {author} {\bibfnamefont {R.~E.}\ \bibnamefont
  {Newnham}},\ }\href@noop {} {\bibfield  {journal} {\bibinfo  {journal} {Acta
  Crystallographica Section A}\ }\textbf {\bibinfo {volume} {54}},\ \bibinfo
  {pages} {729} (\bibinfo {year} {1998})}\BibitemShut {NoStop}%
\bibitem [{\citenamefont {Tokura}(2006)}]{tokura2006}%
  \BibitemOpen
  \bibfield  {author} {\bibinfo {author} {\bibfnamefont {Y.}~\bibnamefont
  {Tokura}},\ }\href@noop {} {\bibfield  {journal} {\bibinfo  {journal}
  {Reports on Progress in Physics}\ }\textbf {\bibinfo {volume} {69}},\
  \bibinfo {pages} {797} (\bibinfo {year} {2006})}\BibitemShut {NoStop}%
\bibitem [{\citenamefont {Lee}\ \emph {et~al.}(2010)\citenamefont {Lee},
  \citenamefont {Fang}, \citenamefont {Vlahos}, \citenamefont {Ke},
  \citenamefont {Jung}, \citenamefont {Kourkoutis}, \citenamefont {Kim},
  \citenamefont {Ryan}, \citenamefont {Heeg}, \citenamefont {Roeckerath} \emph
  {et~al.}}]{lee2010}%
  \BibitemOpen
  \bibfield  {author} {\bibinfo {author} {\bibfnamefont {J.~H.}\ \bibnamefont
  {Lee}}, \bibinfo {author} {\bibfnamefont {L.}~\bibnamefont {Fang}}, \bibinfo
  {author} {\bibfnamefont {E.}~\bibnamefont {Vlahos}}, \bibinfo {author}
  {\bibfnamefont {X.}~\bibnamefont {Ke}}, \bibinfo {author} {\bibfnamefont
  {Y.~W.}\ \bibnamefont {Jung}}, \bibinfo {author} {\bibfnamefont {L.~F.}\
  \bibnamefont {Kourkoutis}}, \bibinfo {author} {\bibfnamefont {J.-W.}\
  \bibnamefont {Kim}}, \bibinfo {author} {\bibfnamefont {P.~J.}\ \bibnamefont
  {Ryan}}, \bibinfo {author} {\bibfnamefont {T.}~\bibnamefont {Heeg}}, \bibinfo
  {author} {\bibfnamefont {M.}~\bibnamefont {Roeckerath}},  \emph {et~al.},\
  }\href@noop {} {\bibfield  {journal} {\bibinfo  {journal} {Nature}\ }\textbf
  {\bibinfo {volume} {466}},\ \bibinfo {pages} {954} (\bibinfo {year}
  {2010})}\BibitemShut {NoStop}%
\bibitem [{\citenamefont {Ryan}\ \emph {et~al.}(2013)\citenamefont {Ryan},
  \citenamefont {Kim}, \citenamefont {Birol}, \citenamefont {Thompson},
  \citenamefont {Lee}, \citenamefont {Ke}, \citenamefont {Normile},
  \citenamefont {Karapetrova}, \citenamefont {Schiffer}, \citenamefont {Brown},
  \citenamefont {Fennie},\ and\ \citenamefont {Schlom}}]{ryan2013}%
  \BibitemOpen
  \bibfield  {author} {\bibinfo {author} {\bibfnamefont {P.}~\bibnamefont
  {Ryan}}, \bibinfo {author} {\bibfnamefont {J.-W.}\ \bibnamefont {Kim}},
  \bibinfo {author} {\bibfnamefont {T.}~\bibnamefont {Birol}}, \bibinfo
  {author} {\bibfnamefont {P.}~\bibnamefont {Thompson}}, \bibinfo {author}
  {\bibfnamefont {J.-H.}\ \bibnamefont {Lee}}, \bibinfo {author} {\bibfnamefont
  {X.}~\bibnamefont {Ke}}, \bibinfo {author} {\bibfnamefont {P.}~\bibnamefont
  {Normile}}, \bibinfo {author} {\bibfnamefont {E.}~\bibnamefont
  {Karapetrova}}, \bibinfo {author} {\bibfnamefont {P.}~\bibnamefont
  {Schiffer}}, \bibinfo {author} {\bibfnamefont {S.}~\bibnamefont {Brown}},
  \bibinfo {author} {\bibfnamefont {C.}~\bibnamefont {Fennie}}, \ and\ \bibinfo
  {author} {\bibfnamefont {D.}~\bibnamefont {Schlom}},\ }\href@noop {}
  {\bibfield  {journal} {\bibinfo  {journal} {Nature Communications}\ }\textbf
  {\bibinfo {volume} {4}},\ \bibinfo {pages} {1334} (\bibinfo {year}
  {2013})}\BibitemShut {NoStop}%
\bibitem [{\citenamefont {Akamatsu}\ \emph {et~al.}(2011)\citenamefont
  {Akamatsu}, \citenamefont {Kumagai}, \citenamefont {Oba}, \citenamefont
  {Fujita}, \citenamefont {Murakami}, \citenamefont {Tanaka},\ and\
  \citenamefont {Tanaka}}]{akamatsu2011}%
  \BibitemOpen
  \bibfield  {author} {\bibinfo {author} {\bibfnamefont {H.}~\bibnamefont
  {Akamatsu}}, \bibinfo {author} {\bibfnamefont {Y.}~\bibnamefont {Kumagai}},
  \bibinfo {author} {\bibfnamefont {F.}~\bibnamefont {Oba}}, \bibinfo {author}
  {\bibfnamefont {K.}~\bibnamefont {Fujita}}, \bibinfo {author} {\bibfnamefont
  {H.}~\bibnamefont {Murakami}}, \bibinfo {author} {\bibfnamefont
  {K.}~\bibnamefont {Tanaka}}, \ and\ \bibinfo {author} {\bibfnamefont
  {I.}~\bibnamefont {Tanaka}},\ }\href@noop {} {\bibfield  {journal} {\bibinfo
  {journal} {Physical Review B}\ }\textbf {\bibinfo {volume} {83}},\ \bibinfo
  {pages} {214421} (\bibinfo {year} {2011})}\BibitemShut {NoStop}%
\bibitem [{\citenamefont {Kolodiazhnyi}\ \emph {et~al.}(2012)\citenamefont
  {Kolodiazhnyi}, \citenamefont {Valant}, \citenamefont {Williams},
  \citenamefont {Bugnet}, \citenamefont {Botton}, \citenamefont {Ohashi},\ and\
  \citenamefont {Sakka}}]{kolodiazhnyi2012}%
  \BibitemOpen
  \bibfield  {author} {\bibinfo {author} {\bibfnamefont {T.}~\bibnamefont
  {Kolodiazhnyi}}, \bibinfo {author} {\bibfnamefont {M.}~\bibnamefont
  {Valant}}, \bibinfo {author} {\bibfnamefont {J.~R.}\ \bibnamefont
  {Williams}}, \bibinfo {author} {\bibfnamefont {M.}~\bibnamefont {Bugnet}},
  \bibinfo {author} {\bibfnamefont {G.~A.}\ \bibnamefont {Botton}}, \bibinfo
  {author} {\bibfnamefont {N.}~\bibnamefont {Ohashi}}, \ and\ \bibinfo {author}
  {\bibfnamefont {Y.}~\bibnamefont {Sakka}},\ }\href@noop {} {\bibfield
  {journal} {\bibinfo  {journal} {Journal of Applied Physics}\ }\textbf
  {\bibinfo {volume} {112}},\ \bibinfo {eid} {083719} (\bibinfo {year}
  {2012})}\BibitemShut {NoStop}%
\bibitem [{\citenamefont {Rushchanskii}\ \emph {et~al.}(2012)\citenamefont
  {Rushchanskii}, \citenamefont {Spaldin},\ and\ \citenamefont
  {Le{\v{z}}ai{\'c}}}]{rushchanskii2012}%
  \BibitemOpen
  \bibfield  {author} {\bibinfo {author} {\bibfnamefont {K.~Z.}\ \bibnamefont
  {Rushchanskii}}, \bibinfo {author} {\bibfnamefont {N.~A.}\ \bibnamefont
  {Spaldin}}, \ and\ \bibinfo {author} {\bibfnamefont {M.}~\bibnamefont
  {Le{\v{z}}ai{\'c}}},\ }\href@noop {} {\bibfield  {journal} {\bibinfo
  {journal} {Physical Review B}\ }\textbf {\bibinfo {volume} {85}},\ \bibinfo
  {pages} {104109} (\bibinfo {year} {2012})}\BibitemShut {NoStop}%
\bibitem [{\citenamefont {Yang}\ \emph {et~al.}(2012)\citenamefont {Yang},
  \citenamefont {Ren}, \citenamefont {Wang},\ and\ \citenamefont
  {Bellaiche}}]{yang2012}%
  \BibitemOpen
  \bibfield  {author} {\bibinfo {author} {\bibfnamefont {Y.}~\bibnamefont
  {Yang}}, \bibinfo {author} {\bibfnamefont {W.}~\bibnamefont {Ren}}, \bibinfo
  {author} {\bibfnamefont {D.}~\bibnamefont {Wang}}, \ and\ \bibinfo {author}
  {\bibfnamefont {L.}~\bibnamefont {Bellaiche}},\ }\href@noop {} {\bibfield
  {journal} {\bibinfo  {journal} {Physical Review Letters}\ }\textbf {\bibinfo
  {volume} {109}},\ \bibinfo {pages} {267602} (\bibinfo {year}
  {2012})}\BibitemShut {NoStop}%
\bibitem [{\citenamefont {Ellis}\ \emph {et~al.}(2012)\citenamefont {Ellis},
  \citenamefont {Uchiyama}, \citenamefont {Tsutsui}, \citenamefont {Sugimoto},
  \citenamefont {Kato}, \citenamefont {Ishikawa},\ and\ \citenamefont
  {Baron}}]{ellis2012}%
  \BibitemOpen
  \bibfield  {author} {\bibinfo {author} {\bibfnamefont {D.~S.}\ \bibnamefont
  {Ellis}}, \bibinfo {author} {\bibfnamefont {H.}~\bibnamefont {Uchiyama}},
  \bibinfo {author} {\bibfnamefont {S.}~\bibnamefont {Tsutsui}}, \bibinfo
  {author} {\bibfnamefont {K.}~\bibnamefont {Sugimoto}}, \bibinfo {author}
  {\bibfnamefont {K.}~\bibnamefont {Kato}}, \bibinfo {author} {\bibfnamefont
  {D.}~\bibnamefont {Ishikawa}}, \ and\ \bibinfo {author} {\bibfnamefont
  {A.~Q.~R.}\ \bibnamefont {Baron}},\ }\href@noop {} {\bibfield  {journal}
  {\bibinfo  {journal} {Phys. Rev. B}\ }\textbf {\bibinfo {volume} {86}},\
  \bibinfo {pages} {220301} (\bibinfo {year} {2012})}\BibitemShut {NoStop}%
\bibitem [{\citenamefont {Fleury}\ \emph {et~al.}(1968)\citenamefont {Fleury},
  \citenamefont {Scott},\ and\ \citenamefont {Worlock}}]{fleury1968}%
  \BibitemOpen
  \bibfield  {author} {\bibinfo {author} {\bibfnamefont {P.~A.}\ \bibnamefont
  {Fleury}}, \bibinfo {author} {\bibfnamefont {J.~F.}\ \bibnamefont {Scott}}, \
  and\ \bibinfo {author} {\bibfnamefont {J.~M.}\ \bibnamefont {Worlock}},\
  }\href@noop {} {\bibfield  {journal} {\bibinfo  {journal} {Phys. Rev. Lett.}\
  }\textbf {\bibinfo {volume} {21}},\ \bibinfo {pages} {16} (\bibinfo {year}
  {1968})}\BibitemShut {NoStop}%
\bibitem [{\citenamefont {Allieta}\ \emph {et~al.}(2012)\citenamefont
  {Allieta}, \citenamefont {Scavini}, \citenamefont {Spalek}, \citenamefont
  {Scagnoli}, \citenamefont {Walker}, \citenamefont {Panagopoulos},
  \citenamefont {Saxena}, \citenamefont {Katsufuji},\ and\ \citenamefont
  {Mazzoli}}]{allieta2012}%
  \BibitemOpen
  \bibfield  {author} {\bibinfo {author} {\bibfnamefont {M.}~\bibnamefont
  {Allieta}}, \bibinfo {author} {\bibfnamefont {M.}~\bibnamefont {Scavini}},
  \bibinfo {author} {\bibfnamefont {L.~J.}\ \bibnamefont {Spalek}}, \bibinfo
  {author} {\bibfnamefont {V.}~\bibnamefont {Scagnoli}}, \bibinfo {author}
  {\bibfnamefont {H.~C.}\ \bibnamefont {Walker}}, \bibinfo {author}
  {\bibfnamefont {C.}~\bibnamefont {Panagopoulos}}, \bibinfo {author}
  {\bibfnamefont {S.~S.}\ \bibnamefont {Saxena}}, \bibinfo {author}
  {\bibfnamefont {T.}~\bibnamefont {Katsufuji}}, \ and\ \bibinfo {author}
  {\bibfnamefont {C.}~\bibnamefont {Mazzoli}},\ }\href@noop {} {\bibfield
  {journal} {\bibinfo  {journal} {Phys. Rev. B}\ }\textbf {\bibinfo {volume}
  {85}},\ \bibinfo {pages} {184107} (\bibinfo {year} {2012})}\BibitemShut
  {NoStop}%
\bibitem [{\citenamefont {M\"uller}\ and\ \citenamefont
  {Burkard}(1979)}]{muller1979}%
  \BibitemOpen
  \bibfield  {author} {\bibinfo {author} {\bibfnamefont {K.~A.}\ \bibnamefont
  {M\"uller}}\ and\ \bibinfo {author} {\bibfnamefont {H.}~\bibnamefont
  {Burkard}},\ }\href@noop {} {\bibfield  {journal} {\bibinfo  {journal} {Phys.
  Rev. B}\ }\textbf {\bibinfo {volume} {19}},\ \bibinfo {pages} {3593}
  (\bibinfo {year} {1979})}\BibitemShut {NoStop}%
\bibitem [{\citenamefont {Weaver}(1959)}]{weaver1959}%
  \BibitemOpen
  \bibfield  {author} {\bibinfo {author} {\bibfnamefont {H.}~\bibnamefont
  {Weaver}},\ }\href@noop {} {\bibfield  {journal} {\bibinfo  {journal}
  {Journal of Physics and Chemistry of Solids}\ }\textbf {\bibinfo {volume}
  {11}},\ \bibinfo {pages} {274 } (\bibinfo {year} {1959})}\BibitemShut
  {NoStop}%
\bibitem [{\citenamefont {Zhong}\ and\ \citenamefont
  {Vanderbilt}(1996)}]{zhong1996}%
  \BibitemOpen
  \bibfield  {author} {\bibinfo {author} {\bibfnamefont {W.}~\bibnamefont
  {Zhong}}\ and\ \bibinfo {author} {\bibfnamefont {D.}~\bibnamefont
  {Vanderbilt}},\ }\href@noop {} {\bibfield  {journal} {\bibinfo  {journal}
  {Phys. Rev. B}\ }\textbf {\bibinfo {volume} {53}},\ \bibinfo {pages} {5047}
  (\bibinfo {year} {1996})}\BibitemShut {NoStop}%
\bibitem [{\citenamefont {Perdew}\ \emph {et~al.}(1996)\citenamefont {Perdew},
  \citenamefont {Burke},\ and\ \citenamefont {Ernzerhof}}]{PBE}%
  \BibitemOpen
  \bibfield  {author} {\bibinfo {author} {\bibfnamefont {J.~P.}\ \bibnamefont
  {Perdew}}, \bibinfo {author} {\bibfnamefont {K.}~\bibnamefont {Burke}}, \
  and\ \bibinfo {author} {\bibfnamefont {M.}~\bibnamefont {Ernzerhof}},\
  }\href@noop {} {\bibfield  {journal} {\bibinfo  {journal} {Phys. Rev. Lett.}\
  }\textbf {\bibinfo {volume} {77}},\ \bibinfo {pages} {3865} (\bibinfo {year}
  {1996})}\BibitemShut {NoStop}%
\bibitem [{\citenamefont {Bl\"ochl}(1994)}]{PAW1}%
  \BibitemOpen
  \bibfield  {author} {\bibinfo {author} {\bibfnamefont {P.~E.}\ \bibnamefont
  {Bl\"ochl}},\ }\href@noop {} {\bibfield  {journal} {\bibinfo  {journal}
  {Phys. Rev. B}\ }\textbf {\bibinfo {volume} {50}},\ \bibinfo {pages} {17953}
  (\bibinfo {year} {1994})}\BibitemShut {NoStop}%
\bibitem [{\citenamefont {Kresse}\ and\ \citenamefont {Joubert}(1999)}]{PAW2}%
  \BibitemOpen
  \bibfield  {author} {\bibinfo {author} {\bibfnamefont {G.}~\bibnamefont
  {Kresse}}\ and\ \bibinfo {author} {\bibfnamefont {D.}~\bibnamefont
  {Joubert}},\ }\href@noop {} {\bibfield  {journal} {\bibinfo  {journal} {Phys.
  Rev. B}\ }\textbf {\bibinfo {volume} {59}},\ \bibinfo {pages} {1758}
  (\bibinfo {year} {1999})}\BibitemShut {NoStop}%
\bibitem [{\citenamefont {Kresse}\ and\ \citenamefont
  {Furthm\"uller}(1996)}]{VASP1}%
  \BibitemOpen
  \bibfield  {author} {\bibinfo {author} {\bibfnamefont {G.}~\bibnamefont
  {Kresse}}\ and\ \bibinfo {author} {\bibfnamefont {J.}~\bibnamefont
  {Furthm\"uller}},\ }\href@noop {} {\bibfield  {journal} {\bibinfo  {journal}
  {Phys. Rev. B}\ }\textbf {\bibinfo {volume} {54}},\ \bibinfo {pages} {11169}
  (\bibinfo {year} {1996})}\BibitemShut {NoStop}%
\bibitem [{\citenamefont {Kresse}\ and\ \citenamefont
  {Furthmuller}(1996)}]{VASP2}%
  \BibitemOpen
  \bibfield  {author} {\bibinfo {author} {\bibfnamefont {G.}~\bibnamefont
  {Kresse}}\ and\ \bibinfo {author} {\bibfnamefont {J.}~\bibnamefont
  {Furthmuller}},\ }\href@noop {} {\bibfield  {journal} {\bibinfo  {journal}
  {Computational Materials Science}\ }\textbf {\bibinfo {volume} {6}},\
  \bibinfo {pages} {15 } (\bibinfo {year} {1996})}\BibitemShut {NoStop}%
\bibitem [{\citenamefont {Anisimov}\ \emph {et~al.}(1991)\citenamefont
  {Anisimov}, \citenamefont {Zaanen},\ and\ \citenamefont
  {Andersen}}]{DFTU:LDAUTYPE1_1}%
  \BibitemOpen
  \bibfield  {author} {\bibinfo {author} {\bibfnamefont {V.~I.}\ \bibnamefont
  {Anisimov}}, \bibinfo {author} {\bibfnamefont {J.}~\bibnamefont {Zaanen}}, \
  and\ \bibinfo {author} {\bibfnamefont {O.~K.}\ \bibnamefont {Andersen}},\
  }\href@noop {} {\bibfield  {journal} {\bibinfo  {journal} {Phys. Rev. B}\
  }\textbf {\bibinfo {volume} {44}},\ \bibinfo {pages} {943} (\bibinfo {year}
  {1991})}\BibitemShut {NoStop}%
\bibitem [{\citenamefont {Liechtenstein}\ \emph {et~al.}(1995)\citenamefont
  {Liechtenstein}, \citenamefont {Anisimov},\ and\ \citenamefont
  {Zaanen}}]{DFTU:LDAUTYPE1_2}%
  \BibitemOpen
  \bibfield  {author} {\bibinfo {author} {\bibfnamefont {A.~I.}\ \bibnamefont
  {Liechtenstein}}, \bibinfo {author} {\bibfnamefont {V.~I.}\ \bibnamefont
  {Anisimov}}, \ and\ \bibinfo {author} {\bibfnamefont {J.}~\bibnamefont
  {Zaanen}},\ }\href@noop {} {\bibfield  {journal} {\bibinfo  {journal} {Phys.
  Rev. B}\ }\textbf {\bibinfo {volume} {52}},\ \bibinfo {pages} {R5467}
  (\bibinfo {year} {1995})}\BibitemShut {NoStop}%
\bibitem [{\citenamefont {Stokes}\ \emph {et~al.}(2007)\citenamefont {Stokes},
  \citenamefont {Hatch},\ and\ \citenamefont {Campbell}}]{isotropy2007}%
  \BibitemOpen
  \bibfield  {author} {\bibinfo {author} {\bibfnamefont {H.}~\bibnamefont
  {Stokes}}, \bibinfo {author} {\bibfnamefont {D.}~\bibnamefont {Hatch}}, \
  and\ \bibinfo {author} {\bibfnamefont {B.}~\bibnamefont {Campbell}},\ }\href
  {http://stokes.byu.edu/isotropy.html} {\enquote {\bibinfo {title}
  {Isotropy},}\ } (\bibinfo {year} {2007})\BibitemShut {NoStop}%
\bibitem [{\citenamefont {Aroyo}\ \emph
  {et~al.}(2006{\natexlab{a}})\citenamefont {Aroyo}, \citenamefont {Kirov},
  \citenamefont {Capillas}, \citenamefont {Perez-Mato},\ and\ \citenamefont
  {Wondratschek}}]{bilbao1}%
  \BibitemOpen
  \bibfield  {author} {\bibinfo {author} {\bibfnamefont {M.~I.}\ \bibnamefont
  {Aroyo}}, \bibinfo {author} {\bibfnamefont {A.}~\bibnamefont {Kirov}},
  \bibinfo {author} {\bibfnamefont {C.}~\bibnamefont {Capillas}}, \bibinfo
  {author} {\bibfnamefont {J.~M.}\ \bibnamefont {Perez-Mato}}, \ and\ \bibinfo
  {author} {\bibfnamefont {H.}~\bibnamefont {Wondratschek}},\ }\href@noop {}
  {\bibfield  {journal} {\bibinfo  {journal} {Acta Crystallographica Section
  A}\ }\textbf {\bibinfo {volume} {62}},\ \bibinfo {pages} {115} (\bibinfo
  {year} {2006}{\natexlab{a}})}\BibitemShut {NoStop}%
\bibitem [{\citenamefont {Aroyo}\ \emph
  {et~al.}(2006{\natexlab{b}})\citenamefont {Aroyo}, \citenamefont
  {Perez-Mato}, \citenamefont {Capillas}, \citenamefont {Kroumova},
  \citenamefont {Ivantchev}, \citenamefont {Madariaga}, \citenamefont {Kirov},\
  and\ \citenamefont {Wondratschek}}]{bilbao2}%
  \BibitemOpen
  \bibfield  {author} {\bibinfo {author} {\bibfnamefont {M.}~\bibnamefont
  {Aroyo}}, \bibinfo {author} {\bibfnamefont {J.}~\bibnamefont {Perez-Mato}},
  \bibinfo {author} {\bibfnamefont {C.}~\bibnamefont {Capillas}}, \bibinfo
  {author} {\bibfnamefont {E.}~\bibnamefont {Kroumova}}, \bibinfo {author}
  {\bibfnamefont {S.}~\bibnamefont {Ivantchev}}, \bibinfo {author}
  {\bibfnamefont {G.}~\bibnamefont {Madariaga}}, \bibinfo {author}
  {\bibfnamefont {A.}~\bibnamefont {Kirov}}, \ and\ \bibinfo {author}
  {\bibfnamefont {H.}~\bibnamefont {Wondratschek}},\ }\href@noop {} {\bibfield
  {journal} {\bibinfo  {journal} {Zeitschrift fur Kristallographie}\ }\textbf
  {\bibinfo {volume} {221}},\ \bibinfo {pages} {15} (\bibinfo {year}
  {2006}{\natexlab{b}})}\BibitemShut {NoStop}%
\bibitem [{\citenamefont {Aroyo}\ \emph {et~al.}(2011)\citenamefont {Aroyo},
  \citenamefont {Perez-Mato}, \citenamefont {Orobengoa}, \citenamefont {Tasci},
  \citenamefont {de~la Flor},\ and\ \citenamefont {Kirov}}]{bilbao3}%
  \BibitemOpen
  \bibfield  {author} {\bibinfo {author} {\bibfnamefont {M.~I.}\ \bibnamefont
  {Aroyo}}, \bibinfo {author} {\bibfnamefont {J.~M.}\ \bibnamefont
  {Perez-Mato}}, \bibinfo {author} {\bibfnamefont {D.}~\bibnamefont
  {Orobengoa}}, \bibinfo {author} {\bibfnamefont {E.}~\bibnamefont {Tasci}},
  \bibinfo {author} {\bibfnamefont {G.}~\bibnamefont {de~la Flor}}, \ and\
  \bibinfo {author} {\bibfnamefont {A.}~\bibnamefont {Kirov}},\ }\href@noop {}
  {\bibfield  {journal} {\bibinfo  {journal} {Bulgarian Chemical
  Communications}\ }\textbf {\bibinfo {volume} {43}},\ \bibinfo {pages} {183}
  (\bibinfo {year} {2011})}\BibitemShut {NoStop}%
\bibitem [{\citenamefont {{Tasci}}\ \emph {et~al.}(2012)\citenamefont
  {{Tasci}}, \citenamefont {{de La Flor}}, \citenamefont {{Orobengoa}},
  \citenamefont {{Capillas}}, \citenamefont {{Perez-Mato}},\ and\ \citenamefont
  {{Aroyo}}}]{bilbao4}%
  \BibitemOpen
  \bibfield  {author} {\bibinfo {author} {\bibfnamefont {E.~S.}\ \bibnamefont
  {{Tasci}}}, \bibinfo {author} {\bibfnamefont {G.}~\bibnamefont {{de La
  Flor}}}, \bibinfo {author} {\bibfnamefont {D.}~\bibnamefont {{Orobengoa}}},
  \bibinfo {author} {\bibfnamefont {C.}~\bibnamefont {{Capillas}}}, \bibinfo
  {author} {\bibfnamefont {J.~M.}\ \bibnamefont {{Perez-Mato}}}, \ and\
  \bibinfo {author} {\bibfnamefont {M.~I.}\ \bibnamefont {{Aroyo}}},\ }in\
  \href@noop {} {\emph {\bibinfo {booktitle} {European Physical Journal Web of
  Conferences}}},\ \bibinfo {series} {European Physical Journal Web of
  Conferences}, Vol.~\bibinfo {volume} {22}\ (\bibinfo {year} {2012})\
  p.~\bibinfo {pages} {9}\BibitemShut {NoStop}%
\bibitem [{\citenamefont {Momma}\ and\ \citenamefont
  {Izumi}(2008)}]{vesta2008}%
  \BibitemOpen
  \bibfield  {author} {\bibinfo {author} {\bibfnamefont {K.}~\bibnamefont
  {Momma}}\ and\ \bibinfo {author} {\bibfnamefont {F.}~\bibnamefont {Izumi}},\
  }\href@noop {} {\bibfield  {journal} {\bibinfo  {journal} {Journal of Applied
  Crystallography}\ }\textbf {\bibinfo {volume} {41}},\ \bibinfo {pages} {653}
  (\bibinfo {year} {2008})}\BibitemShut {NoStop}%
\bibitem [{\citenamefont {Marzari}\ \emph {et~al.}(2012)\citenamefont
  {Marzari}, \citenamefont {Mostofi}, \citenamefont {Yates}, \citenamefont
  {Souza},\ and\ \citenamefont {Vanderbilt}}]{WANNIER_REVIEW}%
  \BibitemOpen
  \bibfield  {author} {\bibinfo {author} {\bibfnamefont {N.}~\bibnamefont
  {Marzari}}, \bibinfo {author} {\bibfnamefont {A.~A.}\ \bibnamefont
  {Mostofi}}, \bibinfo {author} {\bibfnamefont {J.~R.}\ \bibnamefont {Yates}},
  \bibinfo {author} {\bibfnamefont {I.}~\bibnamefont {Souza}}, \ and\ \bibinfo
  {author} {\bibfnamefont {D.}~\bibnamefont {Vanderbilt}},\ }\href@noop {}
  {\bibfield  {journal} {\bibinfo  {journal} {Rev. Mod. Phys.}\ }\textbf
  {\bibinfo {volume} {84}},\ \bibinfo {pages} {1419} (\bibinfo {year}
  {2012})}\BibitemShut {NoStop}%
\bibitem [{\citenamefont {Mostofi}\ \emph {et~al.}(2008)\citenamefont
  {Mostofi}, \citenamefont {Yates}, \citenamefont {Lee}, \citenamefont {Souza},
  \citenamefont {Vanderbilt},\ and\ \citenamefont {Marzari}}]{wannier90}%
  \BibitemOpen
  \bibfield  {author} {\bibinfo {author} {\bibfnamefont {A.~A.}\ \bibnamefont
  {Mostofi}}, \bibinfo {author} {\bibfnamefont {J.~R.}\ \bibnamefont {Yates}},
  \bibinfo {author} {\bibfnamefont {Y.-S.}\ \bibnamefont {Lee}}, \bibinfo
  {author} {\bibfnamefont {I.}~\bibnamefont {Souza}}, \bibinfo {author}
  {\bibfnamefont {D.}~\bibnamefont {Vanderbilt}}, \ and\ \bibinfo {author}
  {\bibfnamefont {N.}~\bibnamefont {Marzari}},\ }\href@noop {} {\bibfield
  {journal} {\bibinfo  {journal} {Computer Physics Communications}\ }\textbf
  {\bibinfo {volume} {178}},\ \bibinfo {pages} {685 } (\bibinfo {year}
  {2008})}\BibitemShut {NoStop}%
\bibitem [{\citenamefont {Akamatsu}\ \emph {et~al.}(2012)\citenamefont
  {Akamatsu}, \citenamefont {Fujita}, \citenamefont {Hayashi}, \citenamefont
  {Kawamoto}, \citenamefont {Kumagai}, \citenamefont {Zong}, \citenamefont
  {Iwata}, \citenamefont {Oba}, \citenamefont {Tanaka},\ and\ \citenamefont
  {Tanaka}}]{akamatsu2012}%
  \BibitemOpen
  \bibfield  {author} {\bibinfo {author} {\bibfnamefont {H.}~\bibnamefont
  {Akamatsu}}, \bibinfo {author} {\bibfnamefont {K.}~\bibnamefont {Fujita}},
  \bibinfo {author} {\bibfnamefont {H.}~\bibnamefont {Hayashi}}, \bibinfo
  {author} {\bibfnamefont {T.}~\bibnamefont {Kawamoto}}, \bibinfo {author}
  {\bibfnamefont {Y.}~\bibnamefont {Kumagai}}, \bibinfo {author} {\bibfnamefont
  {Y.}~\bibnamefont {Zong}}, \bibinfo {author} {\bibfnamefont {K.}~\bibnamefont
  {Iwata}}, \bibinfo {author} {\bibfnamefont {F.}~\bibnamefont {Oba}}, \bibinfo
  {author} {\bibfnamefont {I.}~\bibnamefont {Tanaka}}, \ and\ \bibinfo {author}
  {\bibfnamefont {K.}~\bibnamefont {Tanaka}},\ }\href@noop {} {\bibfield
  {journal} {\bibinfo  {journal} {Inorganic Chemistry}\ }\textbf {\bibinfo
  {volume} {51}},\ \bibinfo {pages} {4560} (\bibinfo {year}
  {2012})}\BibitemShut {NoStop}%
\bibitem [{\citenamefont {Akamatsu}\ \emph {et~al.}(2013)\citenamefont
  {Akamatsu}, \citenamefont {Kumagai}, \citenamefont {Oba}, \citenamefont
  {Fujita}, \citenamefont {Tanaka},\ and\ \citenamefont
  {Tanaka}}]{akamatsu2013}%
  \BibitemOpen
  \bibfield  {author} {\bibinfo {author} {\bibfnamefont {H.}~\bibnamefont
  {Akamatsu}}, \bibinfo {author} {\bibfnamefont {Y.}~\bibnamefont {Kumagai}},
  \bibinfo {author} {\bibfnamefont {F.}~\bibnamefont {Oba}}, \bibinfo {author}
  {\bibfnamefont {K.}~\bibnamefont {Fujita}}, \bibinfo {author} {\bibfnamefont
  {K.}~\bibnamefont {Tanaka}}, \ and\ \bibinfo {author} {\bibfnamefont
  {I.}~\bibnamefont {Tanaka}},\ }\href@noop {} {\bibfield  {journal} {\bibinfo
  {journal} {Advanced Functional Materials}\ }\textbf {\bibinfo {volume}
  {23}},\ \bibinfo {pages} {1864} (\bibinfo {year} {2013})}\BibitemShut
  {NoStop}%
\bibitem [{\citenamefont {Goian}\ \emph {et~al.}(2012)\citenamefont {Goian},
  \citenamefont {Kamba}, \citenamefont {Pacherov\'a}, \citenamefont
  {Drahokoupil}, \citenamefont {Palatinus}, \citenamefont
  {Du\ifmmode~\check{s}\else \v{s}\fi{}ek}, \citenamefont
  {Rohl\'\i\ifmmode~\check{c}\else \v{c}\fi{}ek}, \citenamefont {Savinov},
  \citenamefont {Laufek}, \citenamefont {Schranz}, \citenamefont {Fuith},
  \citenamefont {Kachlik}, \citenamefont {Maca}, \citenamefont {Shkabko},
  \citenamefont {Sagarna}, \citenamefont {Weidenkaff},\ and\ \citenamefont
  {Belik}}]{goian2012}%
  \BibitemOpen
  \bibfield  {author} {\bibinfo {author} {\bibfnamefont {V.}~\bibnamefont
  {Goian}}, \bibinfo {author} {\bibfnamefont {S.}~\bibnamefont {Kamba}},
  \bibinfo {author} {\bibfnamefont {O.}~\bibnamefont {Pacherov\'a}}, \bibinfo
  {author} {\bibfnamefont {J.}~\bibnamefont {Drahokoupil}}, \bibinfo {author}
  {\bibfnamefont {L.}~\bibnamefont {Palatinus}}, \bibinfo {author}
  {\bibfnamefont {M.}~\bibnamefont {Du\ifmmode~\check{s}\else \v{s}\fi{}ek}},
  \bibinfo {author} {\bibfnamefont {J.}~\bibnamefont
  {Rohl\'\i\ifmmode~\check{c}\else \v{c}\fi{}ek}}, \bibinfo {author}
  {\bibfnamefont {M.}~\bibnamefont {Savinov}}, \bibinfo {author} {\bibfnamefont
  {F.}~\bibnamefont {Laufek}}, \bibinfo {author} {\bibfnamefont
  {W.}~\bibnamefont {Schranz}}, \bibinfo {author} {\bibfnamefont
  {A.}~\bibnamefont {Fuith}}, \bibinfo {author} {\bibfnamefont
  {M.}~\bibnamefont {Kachlik}}, \bibinfo {author} {\bibfnamefont
  {K.}~\bibnamefont {Maca}}, \bibinfo {author} {\bibfnamefont {A.}~\bibnamefont
  {Shkabko}}, \bibinfo {author} {\bibfnamefont {L.}~\bibnamefont {Sagarna}},
  \bibinfo {author} {\bibfnamefont {A.}~\bibnamefont {Weidenkaff}}, \ and\
  \bibinfo {author} {\bibfnamefont {A.~A.}\ \bibnamefont {Belik}},\ }\href@noop
  {} {\bibfield  {journal} {\bibinfo  {journal} {Phys. Rev. B}\ }\textbf
  {\bibinfo {volume} {86}},\ \bibinfo {pages} {054112} (\bibinfo {year}
  {2012})}\BibitemShut {NoStop}%
\bibitem [{Note1()}]{Note1}%
  \BibitemOpen
  \bibinfo {note} {It is pleasing that the original DFT paper determined a
  value of U similar to this by comparing the calculated magnetic exchange
  parameters to experiment, and then predicted a $\omega _{SM}$ frequency
  remarkably close to the experiment.}\BibitemShut {Stop}%
\bibitem [{\citenamefont {Cochran}\ and\ \citenamefont
  {Cowley}(1962)}]{cochran1962}%
  \BibitemOpen
  \bibfield  {author} {\bibinfo {author} {\bibfnamefont {W.}~\bibnamefont
  {Cochran}}\ and\ \bibinfo {author} {\bibfnamefont {R.~A.}\ \bibnamefont
  {Cowley}},\ }\href@noop {} {\bibfield  {journal} {\bibinfo  {journal}
  {Journal of Physics and Chemistry of Solids}\ }\textbf {\bibinfo {volume}
  {23}},\ \bibinfo {pages} {447} (\bibinfo {year} {1962})}\BibitemShut
  {NoStop}%
\bibitem [{\citenamefont {Bersuker}\ and\ \citenamefont
  {Vekhter}(1978)}]{bersuker1978}%
  \BibitemOpen
  \bibfield  {author} {\bibinfo {author} {\bibfnamefont {I.~B.}\ \bibnamefont
  {Bersuker}}\ and\ \bibinfo {author} {\bibfnamefont {B.~G.}\ \bibnamefont
  {Vekhter}},\ }\href@noop {} {\bibfield  {journal} {\bibinfo  {journal}
  {Ferroelectrics}\ }\textbf {\bibinfo {volume} {19}},\ \bibinfo {pages} {137}
  (\bibinfo {year} {1978})}\BibitemShut {NoStop}%
\bibitem [{\citenamefont {Cohen}(1992)}]{cohen1992}%
  \BibitemOpen
  \bibfield  {author} {\bibinfo {author} {\bibfnamefont {R.~E.}\ \bibnamefont
  {Cohen}},\ }\href@noop {} {\bibfield  {journal} {\bibinfo  {journal}
  {Nature}\ }\textbf {\bibinfo {volume} {358}},\ \bibinfo {pages} {136}
  (\bibinfo {year} {1992})}\BibitemShut {NoStop}%
\bibitem [{\citenamefont {Rabe}\ \emph {et~al.}(2007)\citenamefont {Rabe},
  \citenamefont {Ahn},\ and\ \citenamefont {Triscone}}]{rabe2007}%
  \BibitemOpen
  \bibfield  {author} {\bibinfo {author} {\bibfnamefont {K.}~\bibnamefont
  {Rabe}}, \bibinfo {author} {\bibfnamefont {C.}~\bibnamefont {Ahn}}, \ and\
  \bibinfo {author} {\bibfnamefont {J.}~\bibnamefont {Triscone}},\ }\href@noop
  {} {\emph {\bibinfo {title} {Physics of Ferroelectrics: A Modern
  Perspective}}},\ Topics in applied physics\ (\bibinfo  {publisher}
  {Springer-Verlag Berlin/Heidelberg},\ \bibinfo {year} {2007})\BibitemShut
  {NoStop}%
\bibitem [{\citenamefont {Bersuker}(2012)}]{bersuker2012}%
  \BibitemOpen
  \bibfield  {author} {\bibinfo {author} {\bibfnamefont {I.~B.}\ \bibnamefont
  {Bersuker}},\ }\href@noop {} {\bibfield  {journal} {\bibinfo  {journal}
  {Phys. Rev. Lett.}\ }\textbf {\bibinfo {volume} {108}},\ \bibinfo {pages}
  {137202} (\bibinfo {year} {2012})}\BibitemShut {NoStop}%
\bibitem [{Note2()}]{Note2}%
  \BibitemOpen
  \bibinfo {note} {All of the Wannier functions presented in this study are
  calculated using an energy range that covers only the Eu f bands. No
  unentanglement is required since these bands are well separated in energy
  from others.}\BibitemShut {Stop}%
\bibitem [{\citenamefont {Ranjan}\ \emph {et~al.}(2009)\citenamefont {Ranjan},
  \citenamefont {Nabi},\ and\ \citenamefont {Pentcheva}}]{ranjan2009}%
  \BibitemOpen
  \bibfield  {author} {\bibinfo {author} {\bibfnamefont {R.}~\bibnamefont
  {Ranjan}}, \bibinfo {author} {\bibfnamefont {H.~S.}\ \bibnamefont {Nabi}}, \
  and\ \bibinfo {author} {\bibfnamefont {R.}~\bibnamefont {Pentcheva}},\
  }\href@noop {} {\bibfield  {journal} {\bibinfo  {journal} {Journal of Applied
  Physics}\ }\textbf {\bibinfo {volume} {105}},\ \bibinfo {pages} {053905}
  (\bibinfo {year} {2009})}\BibitemShut {NoStop}%
\bibitem [{Note3()}]{Note3}%
  \BibitemOpen
  \bibinfo {note} {While the exact quantitative value of $\sigma _{Ti}$ depends
  on the details of the procedure used to calculate the site projected DOS,
  such as the radius of the spheres used; the two trends that we report are
  robust against changes in the sphere size.}\BibitemShut {Stop}%
\bibitem [{\citenamefont {Singh}\ \emph {et~al.}(2006)\citenamefont {Singh},
  \citenamefont {Ghita}, \citenamefont {Fornari},\ and\ \citenamefont
  {Halilov}}]{singh2006}%
  \BibitemOpen
  \bibfield  {author} {\bibinfo {author} {\bibfnamefont {D.~J.}\ \bibnamefont
  {Singh}}, \bibinfo {author} {\bibfnamefont {M.}~\bibnamefont {Ghita}},
  \bibinfo {author} {\bibfnamefont {M.}~\bibnamefont {Fornari}}, \ and\
  \bibinfo {author} {\bibfnamefont {S.~V.}\ \bibnamefont {Halilov}},\
  }\href@noop {} {\bibfield  {journal} {\bibinfo  {journal} {Ferroelectrics}\
  }\textbf {\bibinfo {volume} {338}},\ \bibinfo {pages} {{1489+}} (\bibinfo
  {year} {2006})}\BibitemShut {NoStop}%
\bibitem [{\citenamefont {Khomskii}(2006)}]{khomskii2006}%
  \BibitemOpen
  \bibfield  {author} {\bibinfo {author} {\bibfnamefont {D.}~\bibnamefont
  {Khomskii}},\ }\href@noop {} {\bibfield  {journal} {\bibinfo  {journal}
  {Journal of Magnetism and Magnetic Materials}\ }\textbf {\bibinfo {volume}
  {306}},\ \bibinfo {pages} {1} (\bibinfo {year} {2006})}\BibitemShut {NoStop}%
\bibitem [{\citenamefont {Hill}(2000)}]{hill2000}%
  \BibitemOpen
  \bibfield  {author} {\bibinfo {author} {\bibfnamefont {N.}~\bibnamefont
  {Hill}},\ }\href@noop {} {\bibfield  {journal} {\bibinfo  {journal} {Journal
  of Phsical Chemistry B}\ }\textbf {\bibinfo {volume} {104}},\ \bibinfo
  {pages} {6694} (\bibinfo {year} {2000})}\BibitemShut {NoStop}%
\bibitem [{Note4()}]{Note4}%
  \BibitemOpen
  \bibinfo {note} {Note that DFT with the LDA or GGA approximations is
  essentially a mean field theory \cite {DFTU:LDAUTYPE1_2} and as a result such
  correlated processes are not included in it. However, the requirement that
  the Kohn-Sham states (and the corresponding Wannier states) are orthonormal
  essentially leads to the same result that if the Eu spins are antiparallel
  the $f$ electrons can delocalize to the $d$ states of the Ti ion
  more.}\BibitemShut {Stop}%
\bibitem [{\citenamefont {Haeni}\ \emph {et~al.}(2004)\citenamefont {Haeni},
  \citenamefont {Irvin}, \citenamefont {Chang}, \citenamefont {Uecker},
  \citenamefont {Reiche}, \citenamefont {Li}, \citenamefont {Choudhury},
  \citenamefont {Tian}, \citenamefont {Hawley}, \citenamefont {Craigo} \emph
  {et~al.}}]{haeni2004}%
  \BibitemOpen
  \bibfield  {author} {\bibinfo {author} {\bibfnamefont {J.}~\bibnamefont
  {Haeni}}, \bibinfo {author} {\bibfnamefont {P.}~\bibnamefont {Irvin}},
  \bibinfo {author} {\bibfnamefont {W.}~\bibnamefont {Chang}}, \bibinfo
  {author} {\bibfnamefont {R.}~\bibnamefont {Uecker}}, \bibinfo {author}
  {\bibfnamefont {P.}~\bibnamefont {Reiche}}, \bibinfo {author} {\bibfnamefont
  {Y.}~\bibnamefont {Li}}, \bibinfo {author} {\bibfnamefont {S.}~\bibnamefont
  {Choudhury}}, \bibinfo {author} {\bibfnamefont {W.}~\bibnamefont {Tian}},
  \bibinfo {author} {\bibfnamefont {M.}~\bibnamefont {Hawley}}, \bibinfo
  {author} {\bibfnamefont {B.}~\bibnamefont {Craigo}},  \emph {et~al.},\
  }\href@noop {} {\bibfield  {journal} {\bibinfo  {journal} {Nature}\ }\textbf
  {\bibinfo {volume} {430}},\ \bibinfo {pages} {758} (\bibinfo {year}
  {2004})}\BibitemShut {NoStop}%
\bibitem [{\citenamefont {Bhattacharjee}\ \emph {et~al.}(2009)\citenamefont
  {Bhattacharjee}, \citenamefont {Bousquet},\ and\ \citenamefont
  {Ghosez}}]{bhattacharjee2009}%
  \BibitemOpen
  \bibfield  {author} {\bibinfo {author} {\bibfnamefont {S.}~\bibnamefont
  {Bhattacharjee}}, \bibinfo {author} {\bibfnamefont {E.}~\bibnamefont
  {Bousquet}}, \ and\ \bibinfo {author} {\bibfnamefont {P.}~\bibnamefont
  {Ghosez}},\ }\href@noop {} {\bibfield  {journal} {\bibinfo  {journal} {Phys.
  Rev. Lett.}\ }\textbf {\bibinfo {volume} {102}},\ \bibinfo {pages} {117602}
  (\bibinfo {year} {2009})}\BibitemShut {NoStop}%
\bibitem [{Note5()}]{Note5}%
  \BibitemOpen
  \bibinfo {note} {Note, however, that the data obtained from the calculations
  in the AFM state (red) have a slightly larger slope than the one obtained
  from calculations in the FM state (blue). This indicates that while there are
  other contributions to spin-phonon coupling apart from the mechanism
  discussed in this study, they are relatively small.}\BibitemShut {Stop}%
\bibitem [{\citenamefont {Kolodiazhnyi}\ \emph {et~al.}(2010)\citenamefont
  {Kolodiazhnyi}, \citenamefont {Fujita}, \citenamefont {Wang}, \citenamefont
  {Zong}, \citenamefont {Tanaka}, \citenamefont {Sakka},\ and\ \citenamefont
  {Takayama-Muromachi}}]{kolodiazhnyi2010}%
  \BibitemOpen
  \bibfield  {author} {\bibinfo {author} {\bibfnamefont {T.}~\bibnamefont
  {Kolodiazhnyi}}, \bibinfo {author} {\bibfnamefont {K.}~\bibnamefont
  {Fujita}}, \bibinfo {author} {\bibfnamefont {L.}~\bibnamefont {Wang}},
  \bibinfo {author} {\bibfnamefont {Y.}~\bibnamefont {Zong}}, \bibinfo {author}
  {\bibfnamefont {K.}~\bibnamefont {Tanaka}}, \bibinfo {author} {\bibfnamefont
  {Y.}~\bibnamefont {Sakka}}, \ and\ \bibinfo {author} {\bibfnamefont
  {E.}~\bibnamefont {Takayama-Muromachi}},\ }\href@noop {} {\bibfield
  {journal} {\bibinfo  {journal} {{Applied Physics Letters}}\ }\textbf
  {\bibinfo {volume} {{96}}} (\bibinfo {year} {{2010}})}\BibitemShut {NoStop}%
\bibitem [{\citenamefont {Sai}\ and\ \citenamefont
  {Vanderbilt}(2000)}]{sai2000}%
  \BibitemOpen
  \bibfield  {author} {\bibinfo {author} {\bibfnamefont {N.}~\bibnamefont
  {Sai}}\ and\ \bibinfo {author} {\bibfnamefont {D.}~\bibnamefont
  {Vanderbilt}},\ }\href@noop {} {\bibfield  {journal} {\bibinfo  {journal}
  {Phys. Rev. B}\ }\textbf {\bibinfo {volume} {62}},\ \bibinfo {pages} {13942}
  (\bibinfo {year} {2000})}\BibitemShut {NoStop}%
\bibitem [{\citenamefont {Woodward}(1997)}]{woodward1997}%
  \BibitemOpen
  \bibfield  {author} {\bibinfo {author} {\bibfnamefont {P.~M.}\ \bibnamefont
  {Woodward}},\ }\href@noop {} {\bibfield  {journal} {\bibinfo  {journal} {Acta
  Crystallographica Section B}\ }\textbf {\bibinfo {volume} {53}},\ \bibinfo
  {pages} {44} (\bibinfo {year} {1997})}\BibitemShut {NoStop}%
\bibitem [{Note6()}]{Note6}%
  \BibitemOpen
  \bibinfo {note} {Note that the octahedral rotation angles obtained within DFT
  are roughly 3-4 degrees larger than the experimental value for both EuTiO3
  and SrTiO3. This overestimation with respect to experiment within DFT is
  well-known to occur (see Ref. \protect \rev@citealpnum {sai2000,wahl2008}).
  Recent high-resolution powder diffraction data shows that in EuTiO3, the
  local rotation angle is much larger than the average one.\cite {allieta2012}
  The local value, which is about 8 degrees, agrees well with DFT.\cite
  {yang2012}}\BibitemShut {NoStop}%
\bibitem [{\citenamefont {Stroppa}\ \emph {et~al.}(2010)\citenamefont
  {Stroppa}, \citenamefont {Marsman}, \citenamefont {Kresse},\ and\
  \citenamefont {Picozzi}}]{stroppa2010}%
  \BibitemOpen
  \bibfield  {author} {\bibinfo {author} {\bibfnamefont {A.}~\bibnamefont
  {Stroppa}}, \bibinfo {author} {\bibfnamefont {M.}~\bibnamefont {Marsman}},
  \bibinfo {author} {\bibfnamefont {G.}~\bibnamefont {Kresse}}, \ and\ \bibinfo
  {author} {\bibfnamefont {S.}~\bibnamefont {Picozzi}},\ }\href@noop {}
  {\bibfield  {journal} {\bibinfo  {journal} {New Journal of Physics}\ }\textbf
  {\bibinfo {volume} {12}},\ \bibinfo {pages} {093026} (\bibinfo {year}
  {2010})}\BibitemShut {NoStop}%
\bibitem [{\citenamefont {Wahl}\ \emph {et~al.}(2008)\citenamefont {Wahl},
  \citenamefont {Vogtenhuber},\ and\ \citenamefont {Kresse}}]{wahl2008}%
  \BibitemOpen
  \bibfield  {author} {\bibinfo {author} {\bibfnamefont {R.}~\bibnamefont
  {Wahl}}, \bibinfo {author} {\bibfnamefont {D.}~\bibnamefont {Vogtenhuber}}, \
  and\ \bibinfo {author} {\bibfnamefont {G.}~\bibnamefont {Kresse}},\ }\href
  {\doibase 10.1103/PhysRevB.78.104116} {\bibfield  {journal} {\bibinfo
  {journal} {Phys. Rev. B}\ }\textbf {\bibinfo {volume} {78}},\ \bibinfo
  {pages} {104116} (\bibinfo {year} {2008})}\BibitemShut {NoStop}%
\end{thebibliography}
\end{document}